\documentclass[usenatbib]{mn2e}
\usepackage[psamsfonts]{amssymb}
\usepackage[final]{graphics}

\input psfig.sty

\def\msun{{\rm M}_{\odot}}

\def\trec{{t_{\rm rec}}}
\def\tob{{t_{\rm ob}}}

\def\mdot{{\dot{M}}}

\def\msun{{\rm M_{\odot}}}

\def\etal{{et al.}}

\def\Sig{{\it \Sigma}}

\def\nue{{\it \nu}}
\def\Lambd{{\it \Lambda}}

\author[O.~M. Matthews, R. Speith, G.~A. Wynn \& R.~G. West]{O.~M. Matthews$^1$\footnotemark[1], R. Speith$^2$, G.~A. Wynn$^3$ and R.~G. West$^3$\\
$^1$ Laboratory for Astrophysics, Paul Scherrer Institut, Wurenlingen and Villigen, CH-5232 Villigen PSI, Switzerland\\
$^2$ Institut f{\"u}r Astronomie und Astrophysik, Universit{\"a}t T{\"u}bingen,
Auf der Morgenstelle 10, D-72076 T{\"u}bingen, Germany\\
$^3$ Department of Physics \& Astronomy, University of Leicester, University Road, Leicester, LE1 7RH \\}
\title[Outbursts of WZ Sagittae]{Magnetically moderated outbursts of WZ Sagittae}

\begin{document}
\maketitle
\label{firstpage}

\begin{abstract}
We argue that the quiescent value of the viscosity parameter 
of the accretion disc in WZ~Sge may be 
$\alpha_{\rm{cold}}\sim\rm{0.01}$, in agreement 
with estimates of $\alpha_{\rm{cold}}$ for other dwarf novae.
Assuming the white dwarf in WZ~Sge to be magnetic,
we show that, in quiescence, material close to the white dwarf
can be propelled to larger radii, depleting the inner accretion
disc. The propeller therefore has the effect of stabilizing the
inner disc and allowing the outer disc to accumulate mass.
The outbursts of WZ~Sge are then regulated by the (magnetically
determined) evolution of the surface density of the outer disc
at a radius close to the tidal limit. 
Numerical models confirm that the recurrence time can be significantly
extended in this way. The outbursts are expected to be superoutbursts since the
outer disc radius is forced to exceed the tidal (3:1 resonance) radius.
The large, quiescent disc is expected to be massive, and
to be able to supply the observed mass accretion rate during
outburst. We predict that the long-term spin evolution of the white dwarf 
spin will involve a long cycle of spin up and 
spin down phases. 
\end{abstract}


\footnotetext[1]{email address: owen.matthews@psi.ch}

\section{Introduction}
The SU~UMa stars are a subclass of Dwarf Novae (DN) whose light curves exhibit
superoutburst behaviour. Superoutbursts typically occur at intervals of several
months, interspersed with normal outbursts every few weeks. In general,
superoutbursts are around one magnitude brighter than normal outbursts
and last a few weeks rather than 2-3 days. 
WZ~Sagittae is an unusual SU~UMa star because the recurrence time is 
extremely long and no normal outbursts are observed between superoutbursts.
Superoutbursts occurred in 1913, 1946, 1978 and 2001, with the last of these
coming somewhat earlier than expected after the previously regular
recurrence time $t_{\rm rec} \sim 33$ years.

Normal outbursts in DN can be explained by a thermal-viscous disc
instability due to the partial ionisation of hydrogen \citep[see e.g.][for a review]{can93}. The standard disc instability model (DIM)
assumes that the viscosity of the disc material increases
by around one order of magnitude during outburst, enhancing the
mass-flow rate through the disc. The usual values adopted for the
\citet{sha73} viscosity parameter are 
$\alpha_{\rm cold}\sim\rm{0.01}$ in quiescence and $\alpha_{\rm hot}\sim\rm{0.1}$ in outburst. In an extension to this theory, \citet{osa95} proposed that superoutbursts are due to a thermal-tidal instability. During each normal outburst, only a small fraction of the total disc mass is deposited
onto the white dwarf (WD), leading to an accumulation of disc mass and angular
momentum. The outer disc radius therefore expands until the tidal
radius is reached. At this point, enhanced tidal 
interaction with the secondary star is thought to increase dissipation in the disc and so raise the mass
accretion rate. A superoutburst may then be triggered. 
This model successfully explains the superoutburst behaviour of ordinary SU~UMa
stars using the same viscosity parameters as the standard disc instability
model (DIM).

\citet{sma93} argued that the viscosity in WZ~Sge
must be far lower than the standard values 
(so that $\alpha_{\rm cold}\leq\rm{5}\times\rm{10^{-5}}$), for the following
reasons: \\
\newline
\noindent
{\bf Smak 1} In the standard DIM, the inter-outburst recurrence time is 
governed by the viscous time-scale of the inner accretion disc, which is far shorter than the 
observed value of $\sim 30 \ {\rm yr}$. Lowering $\alpha_{\rm cold}$ slows down the 
viscous evolution of the accretion disc, increasing the inter-outburst
time. \\
\newline
\noindent
{\bf Smak 2} Integrating the critical surface density using the standard DIM
$\alpha_{\rm cold}$ gives a maximum disc mass that is far lower than the
observed estimate of the mass accreted during superoutburst,
$\Delta M_{\rm acc} \sim 10^{24} \ {\rm g}$. Lowering the viscosity
allows more mass to be accumulated in the outer regions of the
accretion disc to fuel the outburst accretion rate. \\
\newline

There is however no obvious reason why the quiescent accretion disc
in WZ Sge should have a much lower viscosity than the other SU~UMa
stars. This has led a number of authors to suggest other reasons to account for the long recurrence time and high disc mass prior to superoutburst. 

In order to produce a long recurrence time without requiring a low
$\alpha_{\rm cold}$ the inner, most unstable regions of the disc must be stabilized. \citet{ham97} and \citet{war96} find that the disc can
be stabilized if the inner regions are removed, either by evaporation
into a coronal layer \citep[see also][]{mey99}, 
or by the presence of a magnetosphere. This need not necessarily lead to an increased recurrence time however. Models by \citet{ham97} suggest a marginally stable disc that requires an episode of enhanced mass
transfer to trigger a disc outburst. The recurrence time in this model
is governed by the mass transfer fluctuation cycle, which has no
obvious physical connection with the regular 33 year outburst cycle 
observed in WZ~Sge. \citet{war96} also find that under
certain conditions the disc will be marginally unstable and produce outside-in
outbursts with the required recurrence time (We explore the case of a marginally stable disc in Section \ref{sec:cod}). However, neither of these models addresses the problem of how to accumulate enough disc mass during
quiescence to explain the mass accreted in outburst {\bf (Smak 2)}.
Only $\sim 10^{21} \ {\rm g}$ is available in the disc just prior to 
outburst so that both of the above models require mass to be added to the 
disc during the
outburst. The authors appeal to irradiation of the secondary star to
increase the transfer rate during outburst and supply the 
missing mass. \citet{war96} only produce
normal outbursts, not superoutbursts in their model. 
The reason for this is that
their disc radius $R_{\rm disc} \sim 1.1 \times 10^{10} \ {\rm cm}$
is far smaller than the tidal radius, so enhanced tidal
interaction with the secondary (required to produce a superoutburst)
is not possible. 

In this paper we argue that the value of $\alpha_{\rm cold}$ 
in WZ Sge may be consistent with standard DIM values ($\alpha_{\rm cold} \approx 0.01$).
In section \ref{sec:qui} we perform a similar calculation to \citet{sma93} 
adopting the system parameters suggested by \citet{spr98}.
 For a fixed disc mass we find that 
$\alpha_{\rm cold}\propto{R_{\rm out}^{\rm 3.9}}$ 
(where $R_{\rm out}$ is the outer disc radius), 
and show that a value of $\alpha_{\rm cold}\approx\rm 0.01$, 
and a moderate increase in ${R_{\rm out}}$ 
results in a value for $\Delta M_{\rm acc}$ consistent with observation.
We then obtain a similar value
for $\alpha_{\rm cold}$ by estimating the cool viscous time-scale from the 
observed superoutburst profile. 

X-ray observations of WZ Sge by \citet{pat98} have revealed a
coherent 27.86 s oscillation in the ASCA 2-6 keV energy band
\citep[see also][]{pat80}. The most convincing explanation for
these data is that WZ Sge contains a magnetic white dwarf.
A paper by \citet{las99} interprets WZ Sge
as a DQ~Her star and the 27.86 s oscillation as the spin period of the
white dwarf ($P_{\rm spin}$).
The authors then suggest that WZ~Sge is an ejector system: i.e. most of the
material transferred from the secondary is ejected from the
system, similar to the case of AE Aqr \citep{wyn97}.
In this model an accretion disc is recreated in outburst, which again
is triggered by a mass transfer event. 

In section \ref{sec:mag} we suggest that if the magnetic torque is not sufficient to 
eject mass completely from the system, but merely propels it further out into the 
Roche Lobe of the WD (we refer to the system as a weak
magnetic propeller), then
{\bf (Smak 1)} and {\bf (Smak 2)} can be explained with a standard DIM 
value for $\alpha_{\rm cold}$. The injection of angular
momentum into the disc by the propeller radically alters the surface
density profile, compared with that of a
conventional DIM disc. The inner disc in such a system is empty, and outbursts 
cannot therefore be triggered there in the normal way. The propeller also inhibits 
accretion, and mass accumulates in the outer disc, increasing the recurrence time significantly. The disc may also be forced to 
expand to the tidal radius (and a little beyond) which could allow further material 
to accumulate in the outer regions of the disc before triggering an outburst. 
This would increase the recurrence time and outburst mass still further, 
solving {\bf (Smak 1)} and {\bf (Smak 2)}. The outburst is initiated as in the 
standard DIM, and does not require any episode of enhanced mass transfer. Simulations 
are performed in one and two dimensions. The former are computationally cheap, so that 
we can present several outburst cycles, but cannot account for the enhanced mass storage 
due to tidally drive disc expansion. 

In Section \ref{sec:obs} we discuss the observational evidence for a truncated accretion disc with reference to observed spin periods and spectral data. Both these results support are in agreement with a truncated disc. In section \ref{sec:spi} we consider the spin evolution of the WD, and whether a spin cycle might exist. Finally, in section \ref{sec:dis} the applications to other CVs are discussed.

\section{The quiescent value of ${\mathbf{\alpha}}$ in WZ Sge}
\label{sec:qui}
\subsection{The quiescent disc mass as a limit on ${\mathbf{\alpha_{\rm cold}}}$}
\label{sec:qdm}
Following \citet{sma93} we estimate a value for $\alpha_{\rm cold}$ by
considering the disc mass immediately prior to outburst. An outburst is triggered
when the surface density at some radius $\Sigma \left( R \right)$ becomes greater than the
maximum value ($\Sigma_{\rm crit}$) allowed on the cool branch of the $\Sigma\ {-
T}$ relation \citep[e.g.][]{can93} at that radius. Therefore, the maximum possible disc mass prior to
outburst is given by:
\begin{equation} \label{int}
M_{\rm max} \simeq \int_{R_{\rm in}}^{R_{\rm out}} {2\pi R \Sigma_{\rm crit} \left( R \right) dR}
\; ,
\end{equation}
where $R_{\rm in}$ and $R_{\rm out}$ are the inner and outer disc radii
respectively. Calculations of the vertical disc structure 
yield \citep[e.g.][]{lud94}
\begin{equation} \label{sig}
\Sigma_{\rm crit}\ \left( R \right) = 670 \ M_{1}^{-0.37} \ R_{10}^{1.10} \
\alpha_{\rm c,0.01}^{-0.8} \ \mu^{0.4} \ {\rm g \ cm^{-2}} \; ,
\end{equation}
where $M_{1}$ is the mass of the WD in solar masses, 
$R_{10}=R/10^{10}$ cm, $\alpha_{\rm{c,0.01}}=\alpha_{\rm cold}/0.01$, and
$\mu$ is the mean molecular weight.
Assuming that $\mu\sim\rm{1}$ and setting $R_{\rm in}=0$, equations (\ref{int}) and (\ref{sig})
can be combined to give
\begin{equation} 
M_{\rm max} \simeq1.36\times{10^{23}} M_{1}^{-0.37} \alpha_{\rm c,0.01}^{-0.8}\ R_{\rm out,10}^{3.1} \; .
\label{mass}
\end{equation}
Although WZ~Sge has a long observational history, the values of some of
its fundamental parameters are still uncertain. For instance, \citet{sma93} deduces that $M_{1} = 0.45$ and $q = M_2/M_1 =
0.13$, whereas the spectroscopic observations of \citet{spr98} (SR hereafter) lead to values of $M_{1} = 1.2$ and $q = 0.075$. \citet{las99} point out that a white dwarf mass of $M_{1} = 0.45$
is too small if it is assumed to be rotating with a period of 27.87 s
(although $M_{1} = 1.2$ is possibly too high), so for the purposes of this paper we shall adopt the system masses suggested by SR. We also adopt an orbital period of $82$ minutes.

\citet{sma93} estimated the total mass 
accreted during outburst to be around 
$\Delta M_{\rm acc} \sim 1$-$2\times 10^{24} \ {\rm g}$.
Setting $\Delta\rm{M_{\rm acc}} \lesssim M_{\rm max}$, and using the
system parameters   $M_{1}=1.2$, 
$q=0.075$ and
$R_{\rm out}=0.37a=1.7\times\rm{10^{10}}cm$ (SR),
we find from equation (\ref{mass}) that $\alpha_{\rm cold}\la\rm{0.006}$.
While this value is around a factor of 2 lower than the standard DIM value of
$\alpha_{\rm cold}\sim\rm{0.01}$, it is 2 orders of magnitude higher
than the value obtained by Smak. 
However, setting $M_{\rm max}\approx\Delta M_{\rm acc}$, we find 
$\alpha_{\rm cold}\propto\ {R_{\rm out}}^{3.9}$, i.e. $\alpha_{\rm cold}$ is 
very sensitive to disc radius. Therefore, any error in the observationally inferred 
value will produce a large discrepancy in $\alpha_{\rm cold}$. The
observational estimate in SR ($R_{\rm out}=0.37a$) was determined from
models of the accretion stream disc impact region, which was found
to be rather extended in the case of WZ~Sge. 
A theoretical estimate of the tidal radius can be found by 
utilizing smoothed particle hydrodynamic calculations 
assuming the system parameters detailed above. This approach yields a tidal radius of 
$R_{\rm out}\simeq\rm{0.5a}$ (see section \ref{sec:threed}). 
Assuming this value for $R_{\rm out}$ and repeating the above calculation
we find that $\alpha_{\rm cold}\la\rm{0.02}$, in good agreement with the 
standard DIM value. We note that the analytic treatment of \citet{osa02} gives a similar $R_{\rm out}$ when the same parameters are used. 

\subsection{Extracting ${\mathbf{\alpha_{\rm cold}}}$ from the outburst lightcurve}
\label{sec:ext}
A second, independent estimate of $\alpha_{\rm cold}$ can be obtained using
the observed outburst profiles \citep{pat81}. Superoutbursts have similar observational profiles  
to the outbursts of x-ray transients \citep{kin98}, 
i.e. they have a rapid rise to outburst
maximum, followed by an exponential decay on the hot viscous
time-scale until the disc is no longer hot enough to remain fully
ionized and falls into quiescence. This produces a steep decline in
brightness on a thermal time-scale followed by a slower decline
as the disc readjusts on the cool viscous time-scale 

\begin{figure}
\begin{center}
\resizebox{80.0mm}{56.5mm}{
\mbox{
\includegraphics{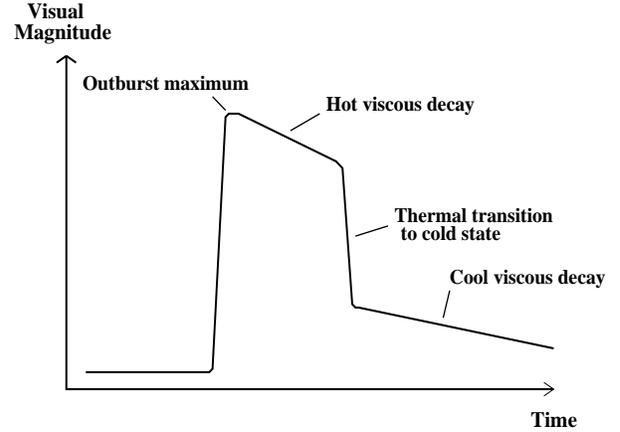}}}
\end{center}
\caption{Schematic diagram of a superoutburst profile.
}
\label{fig:ob}
\end{figure}

\begin{figure}
\begin{center}
\resizebox{80.0mm}{80.0mm}{
\mbox{
\includegraphics{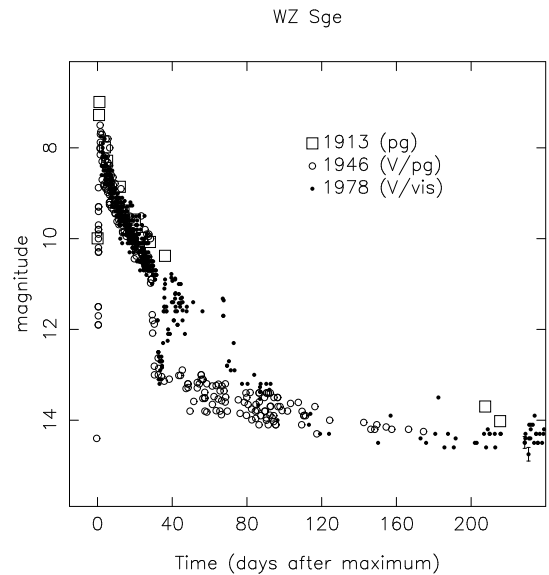}}}
\end{center}
\caption{Plot of optical magnitude during the 1913, 1946 and 1978 outbursts of WZ Sge, from \citet{kul00}. Reprinted with permission from Elsevier.
}
\label{fig:kuulkers}
\end{figure}

Fig \ref{fig:kuulkers} \citep[from][]{kul00} presents  the visual lightcurves for the 1913,
1946 and 1978 outbursts. It is possible to determine $\alpha_{\rm cold}$ by
considering the visual luminosity as
the disc returns to quiescence, i.e. $\sim\rm{30-90\ days}$ after
the outburst began. This is more difficult in the case of the 2001 outburst since the lightcurve is complicated by the presence of echo outbursts. The disc brightness decays exponentially on the
cool viscous time-scale $t_{\rm visc,c}$ (this is clearly seen in the 1946 outburst, but
is less obvious in 1978), such that the
visual luminosity is given by: 
\begin{equation} \label{flux} 
F \left( t \right)=F_{0} \exp \left( \frac{-t}{t_{\rm visc,c}} \right) \; ,
\end{equation} 
where $t$ is time and the cold viscous time-scale is given by
\begin{equation} \label{tvisc}
t_{\rm visc,c} \sim \frac{R^{2}}{\nu_{\rm c}} \; .
\end{equation}
The cold state viscosity is given by \citep{sha73}
\begin{equation}
\label{eqn:shasun}
\nu_{\rm c}\sim\alpha_{\rm cold}{c_{\rm s}H} \; .
\end{equation}
Here $c_s$ is the adiabatic sound speed, and $H$ is the disc scale height. 
During quiescence the disc temperature is $T \sim 5000 \pm 1000 \ {\rm K}$. 
If we assume that $\mu\sim\rm{1}$, the sound speed is then
$c_{\rm s} \simeq \sqrt{\frac{kT}{\mu m_{\rm H}}} \sim 6.4 \times  10^{5} \ {\rm cm \ s^{-1}}$.
We consider the part of the disc from which the majority of the emission should come during the cold viscous decay, namely $R\sim 2 \times 10^{10} \pm 1 \times 10^{10} \ {\rm cm}$. Viscous dissipation is given by 
\begin{equation}
D \left( R \right) = \frac{9}{8} \nu \Sigma \Omega \; , 
\label{eqn:emission}
\end{equation}
where $\Omega$ is the disc angular frequency \citep[e.g.][]{fra01}. The very inner part of the disc is evacuated, while the outer part has a low angular frequency so that our estimate for $R$ seems sensible.
We further assume that $H/R \sim 0.10 \pm 0.05$ for a geometrically thin disc. 
From the 1946 outburst lightcurve it can be seen that, as the disc falls into
quiescence, the visual flux drops by one magnitude in 
$\sim\rm{130}\pm\rm{40}\ \rm{days}$. From equation (\ref{tvisc}) we find
$\alpha_{\rm cold} \sim \rm{0.028\pm\rm{0.022}}$, in reasonable agreement
with the estimate in section \ref{sec:qdm}.

%\begin{figure}
%\begin{center}
%\resizebox{80.0mm}{80.0mm}{
%\mbox{
%\includegraphics{rd6.ps}}}
%\end{center}
%\caption{Steady-state surface density profiles of one-dimensional simulations of the WZ~Sge accretion disc. The solid line is without a stellar magnetic field and dashed line represents a field of $B \sim 1 \ {\rm kG}$. 
%}
%\label{fig:newprofile}
%\end{figure}

\begin{figure}
\begin{center}
\resizebox{88.0mm}{88.0mm}{
\mbox{
\includegraphics{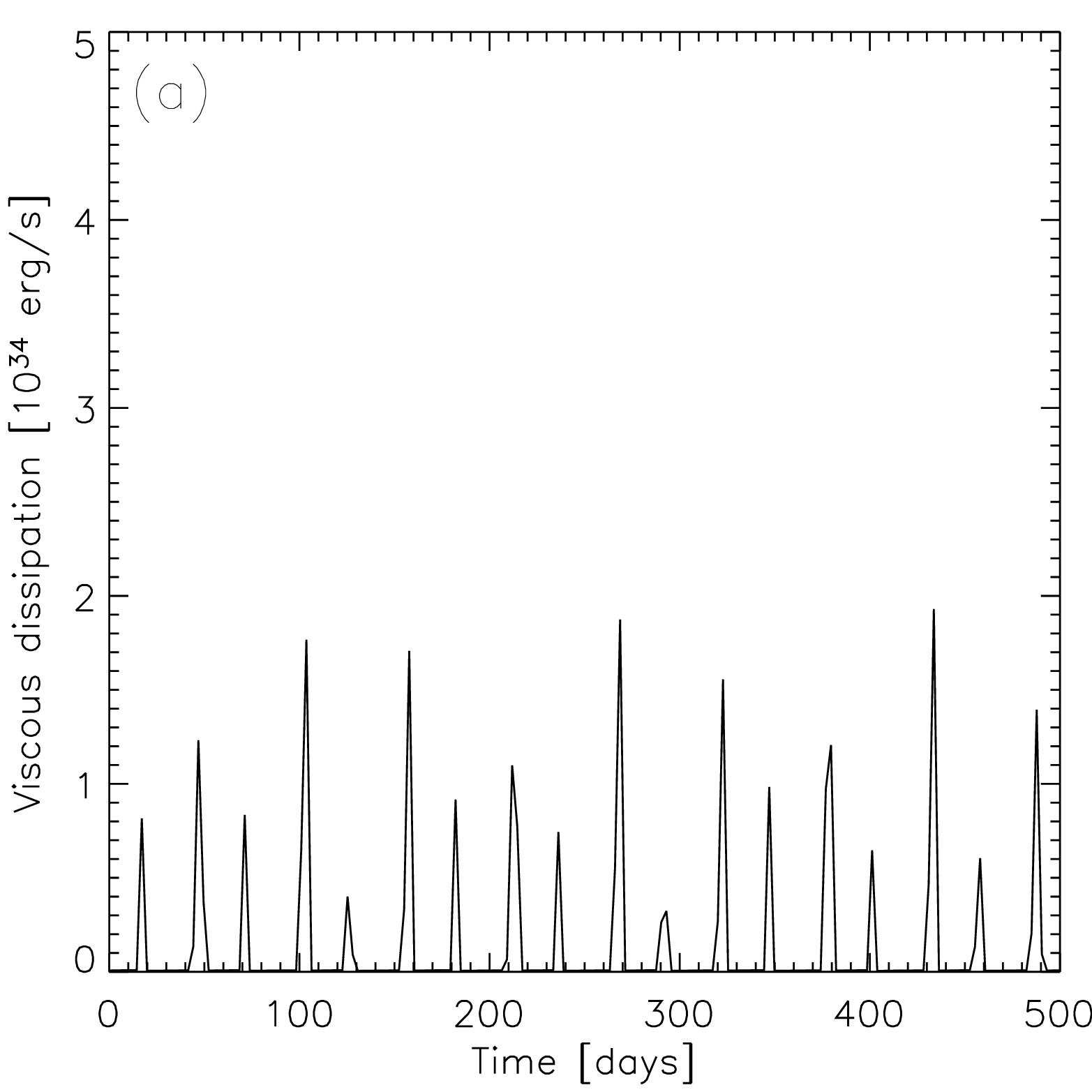}}}
\resizebox{88.0mm}{88.0mm}{
\mbox{
\includegraphics{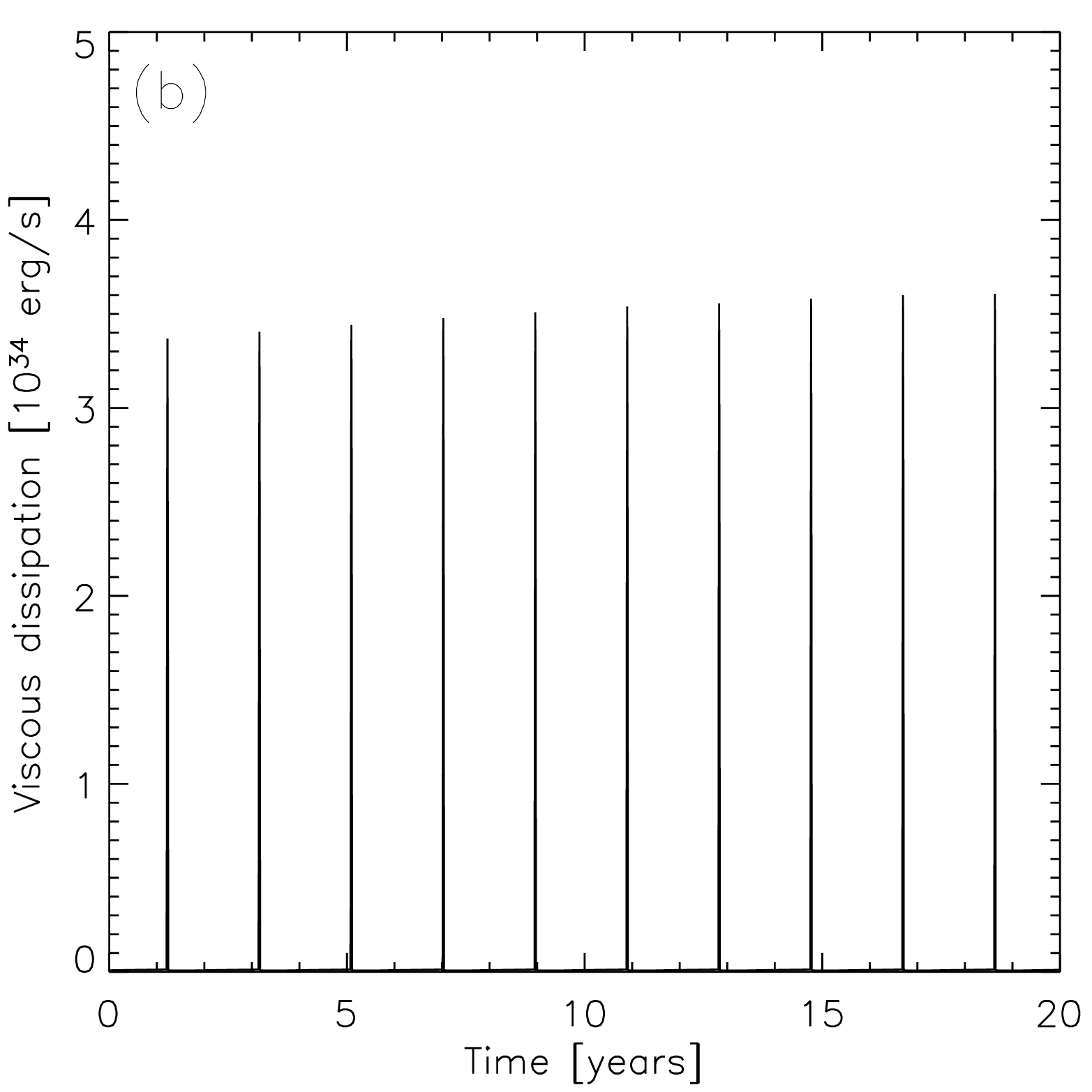}}}
\end{center}
\caption{Plots of dissipation, showing outburst behaviour, in one-dimensional simulations of the WZ~Sge accretion disc. Frame (a) is without a stellar magnetic field and frame (b) frame has a field of $B \sim 1 \ {\rm kG}$.
}
\label{fig:newcurve}
\end{figure}

\begin{figure}
\begin{center}
\resizebox{88.0mm}{88.0mm}{
\mbox{
\includegraphics{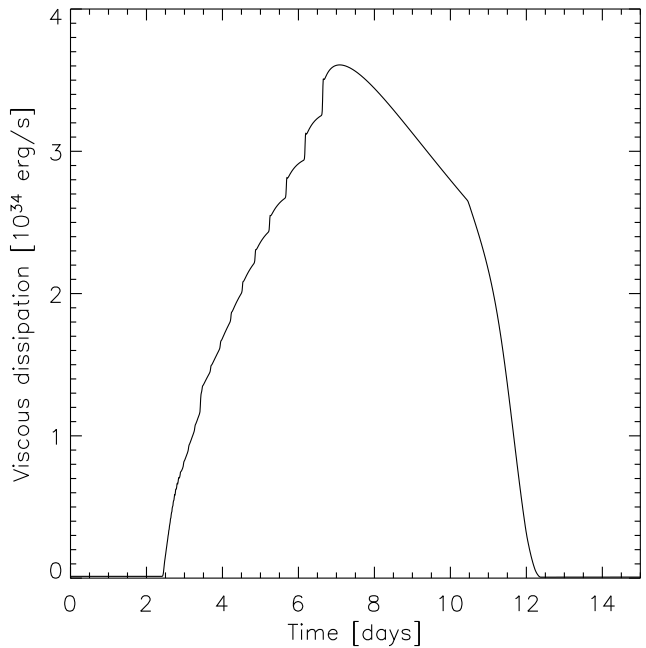}}}
\end{center}
\caption{Plot of dissipation, showing a magnetically moderated outburst in one-dimensional simulations. 
}
\label{fig:obprof}
\end{figure}

\begin{figure}
\begin{center}
\resizebox{80.0mm}{80.0mm}{
\mbox{
\includegraphics{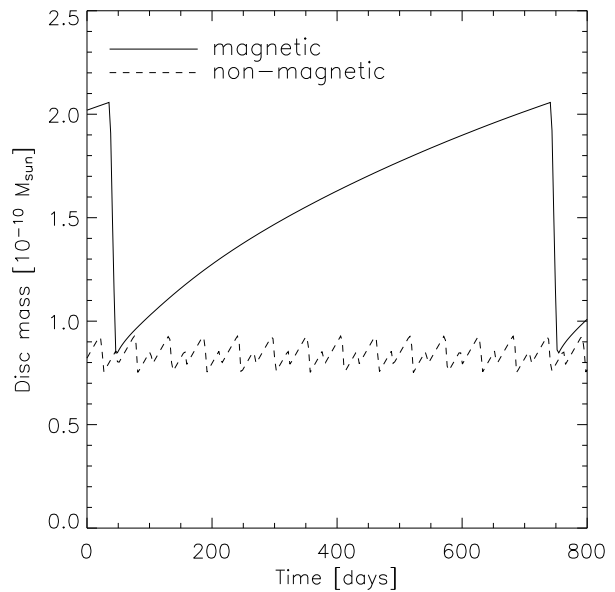}}}
\end{center}
\caption{Plot showing evolution of the accretion disc mass as a function of time magnetic and non-magnetic cases. The magnetic outbursts (where $B \sim 1 \ {\rm kG}$) consume more than half of the accretion disc, about ten times as much as the non-magnetic outbursts. 
}
\label{fig:massevolve}
\end{figure}

\begin{figure}
\begin{center}
\resizebox{80.0mm}{80.0mm}{
\mbox{
\includegraphics{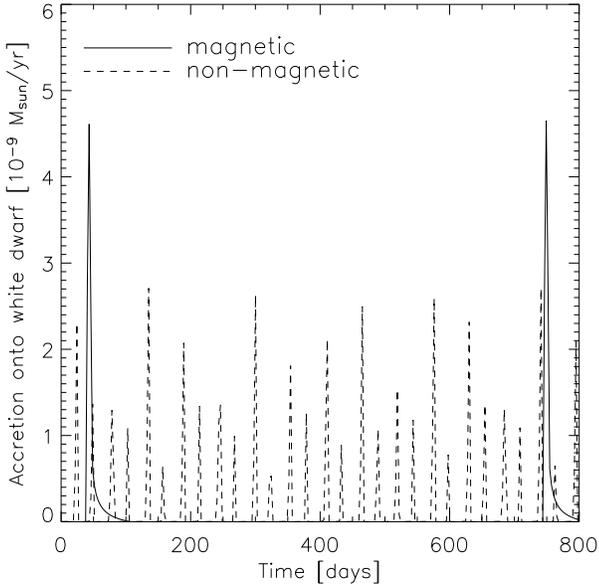}}}
\end{center}
\caption{Plot showing accretion onto the white dwarf as a function of time in magnetic and non-magnetic cases on the same time scale as Fig \ref{fig:massevolve}. The magnetic outbursts have a higher peak accretion rate than the standard outbursts.
}
\label{fig:mdot}
\end{figure}

\begin{figure}
\begin{center}
\resizebox{80.0mm}{80.0mm}{
\mbox{
\includegraphics{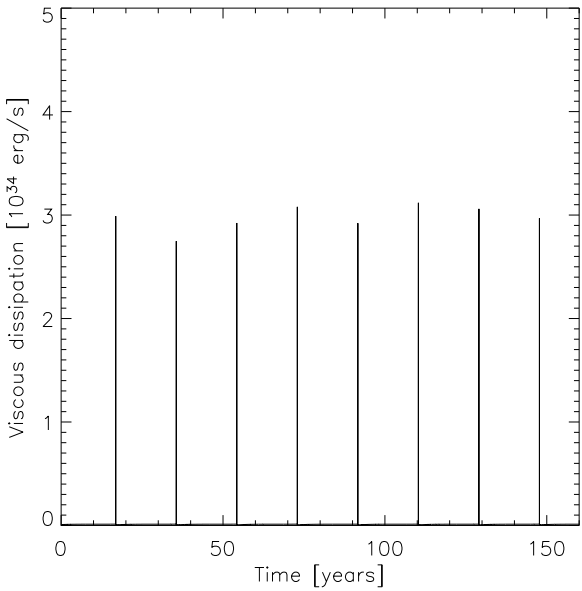}}}
\end{center}
\caption{Plot showing viscous dissipation for the case of a magnetically truncated disc close to marginal stability with $B \sim 1 \ {\rm kG}$. The recurrence time is increased to $t_{\rm rec} \sim 20 \ {\rm yr}$.
}
\label{fig:marginal}
\end{figure}

\begin{figure*}
\begin{center}
\resizebox{88.0mm}{88.0mm}{
\mbox{
\includegraphics{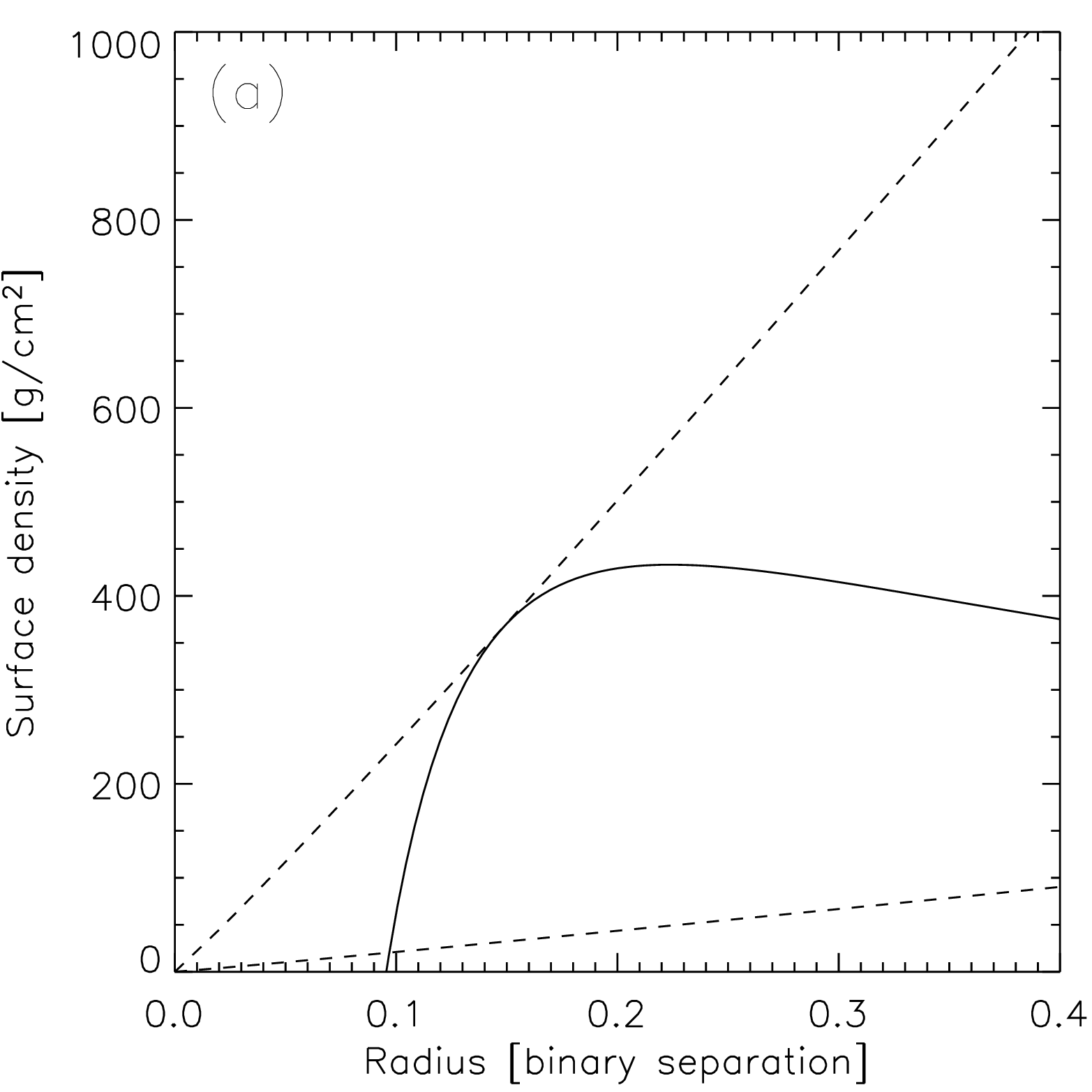}}}
\resizebox{88.0mm}{88.0mm}{
\mbox{
\includegraphics{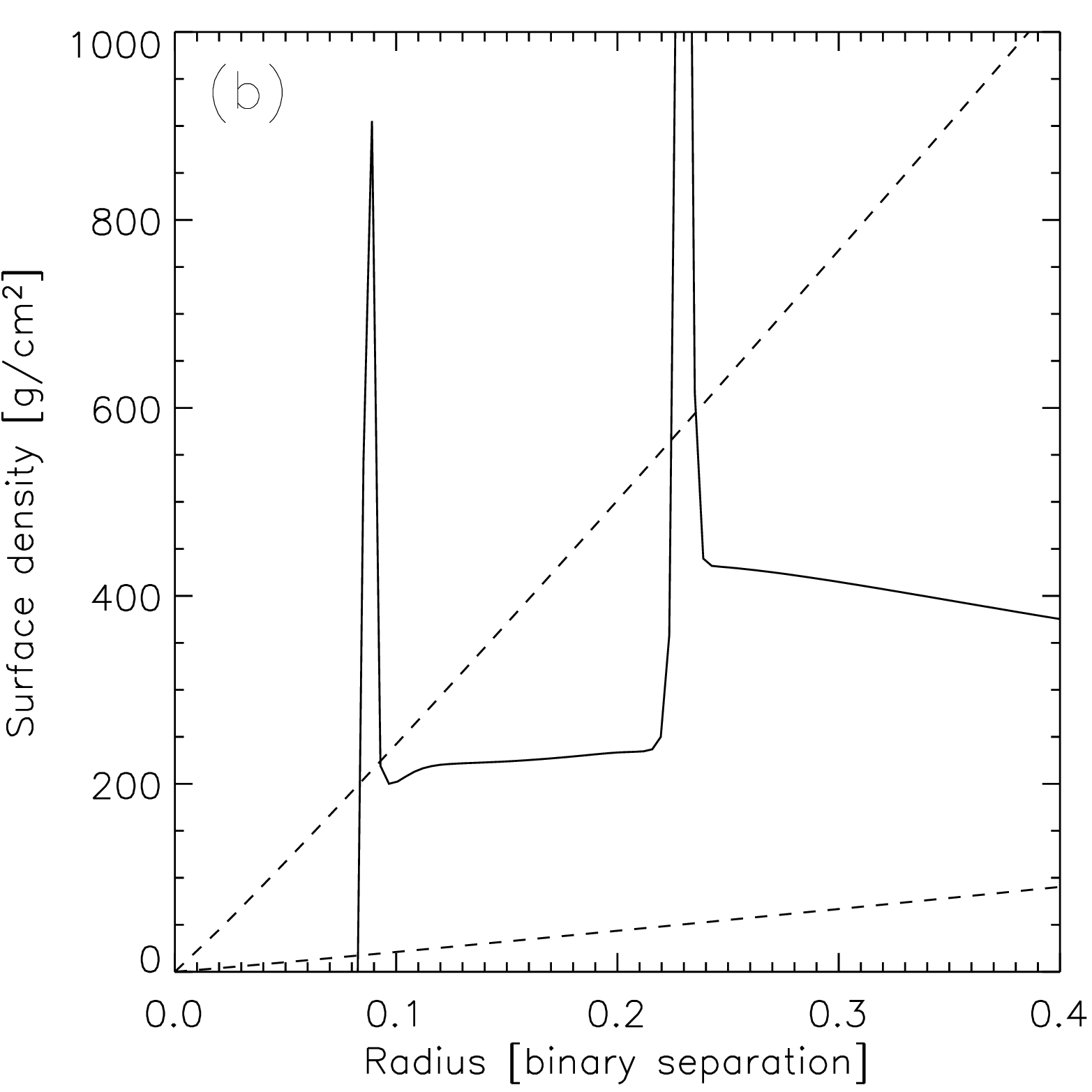}}}
\resizebox{88.0mm}{88.0mm}{
\mbox{
\includegraphics{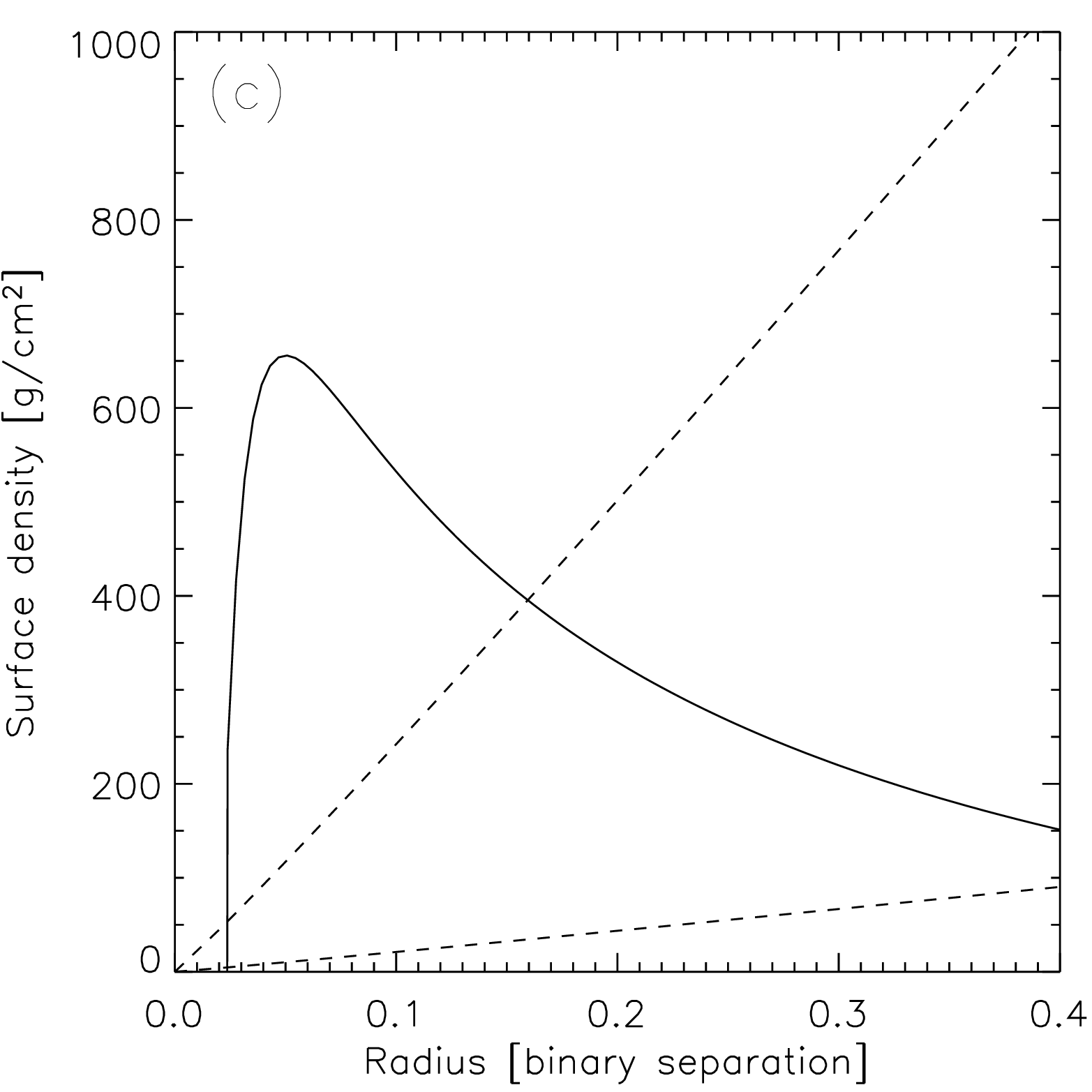}}}
\resizebox{88.0mm}{88.0mm}{
\mbox{
\includegraphics{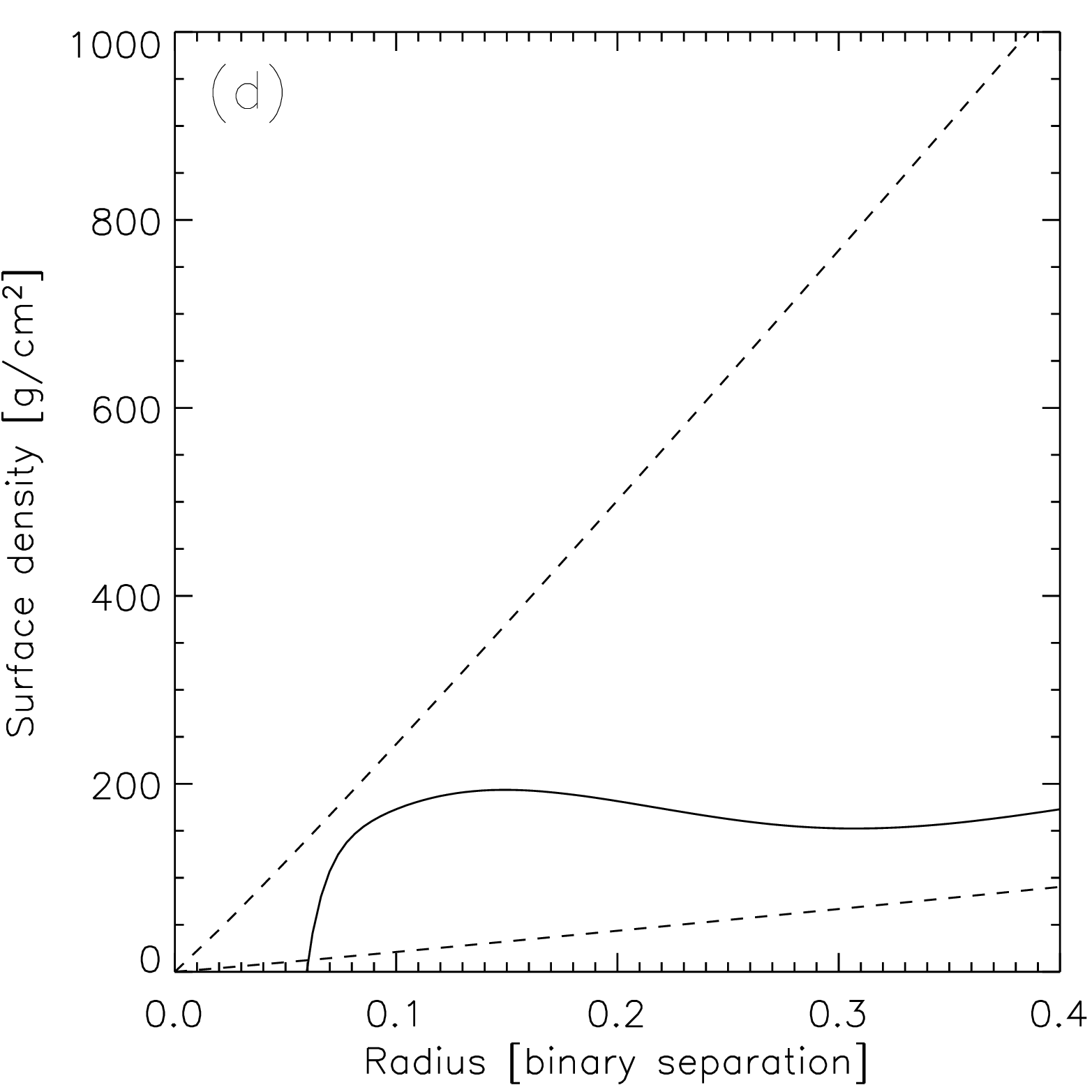}}}
\end{center}
\caption{Surface density profiles of the disc in WZ~Sge as simulated in the one-dimensional code with a magnetic torque applied. Four snapshots are shown in frames (a) to (d). Frame (a) shows the disc just before the onset of an outburst. Frame (b) is seven hours later showing the heating waves propagating inwards and outwards. Frame (c) shows the disc four days later, at the height of the outburst, with the disc in the fully hot state and with no magnetic truncation. Finally, frame (d) shows the early inter-outburst period, eighteen days later, with a greatly depleted disc. The critical trigger densities are plotted with dashed lines.
}
\label{fig:newevol}
\end{figure*}

\begin{figure*}
\begin{center}
\resizebox{55.0mm}{45.0mm}{
\mbox{
\includegraphics{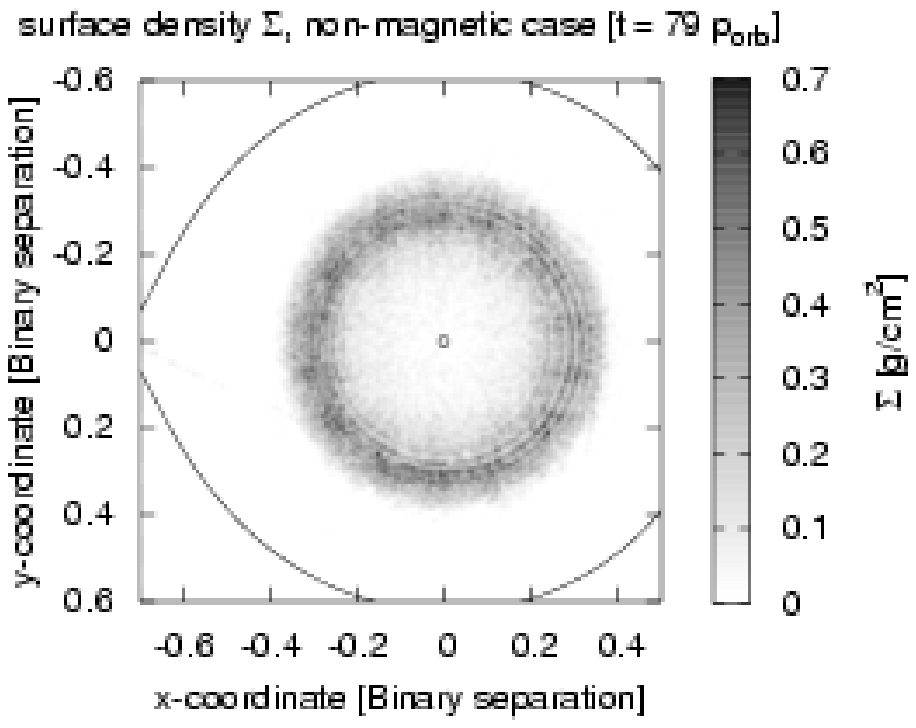}}}
\resizebox{55.0mm}{45.0mm}{
\mbox{
\includegraphics{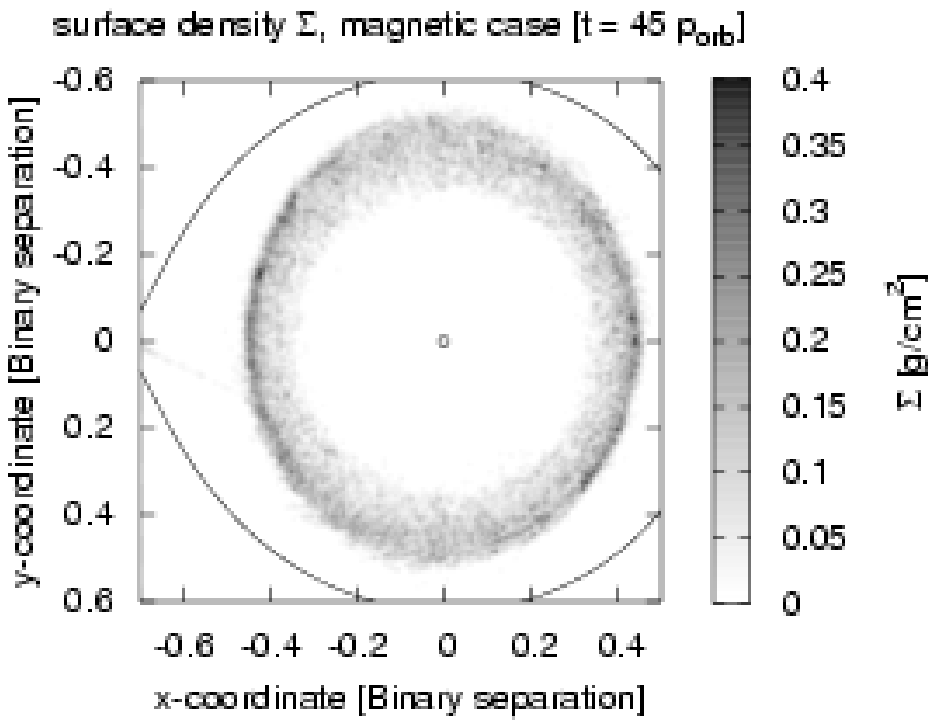}}}
\resizebox{55.0mm}{45.0mm}{
\mbox{
\includegraphics{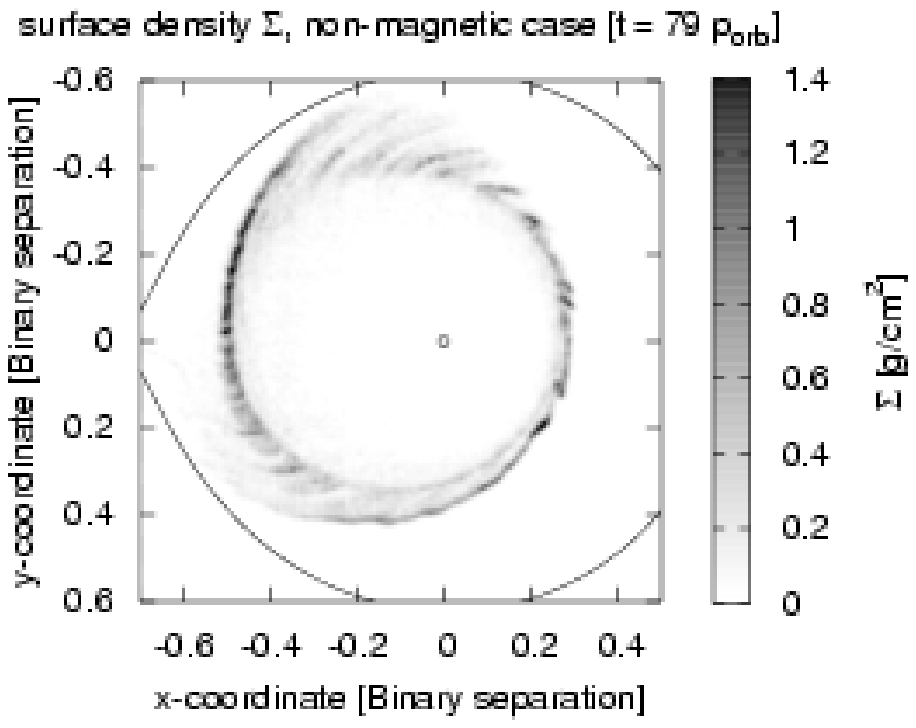}}}
\end{center}
\caption{Grey-scale density plots of the accretion disc in~WZ Sge, simulated in the two-dimensional SPH code. The first plot shows the disc in the non-magnetic case after 79 orbits. The second and third plots show the disc in the magnetic case, after 45 and 79 orbits respectively. The first of these two plots is just before the disc begins to become eccentric and the second shows the later, eccentric disc. The primary Roche lobe is plotted as a solid line and the primary star is marked by a circle (not to scale) in all three plots.
}
\label{fig:disc}
\end{figure*}

\begin{figure*}
\begin{center}
\resizebox{80.0mm}{60.0mm}{
\mbox{
\includegraphics{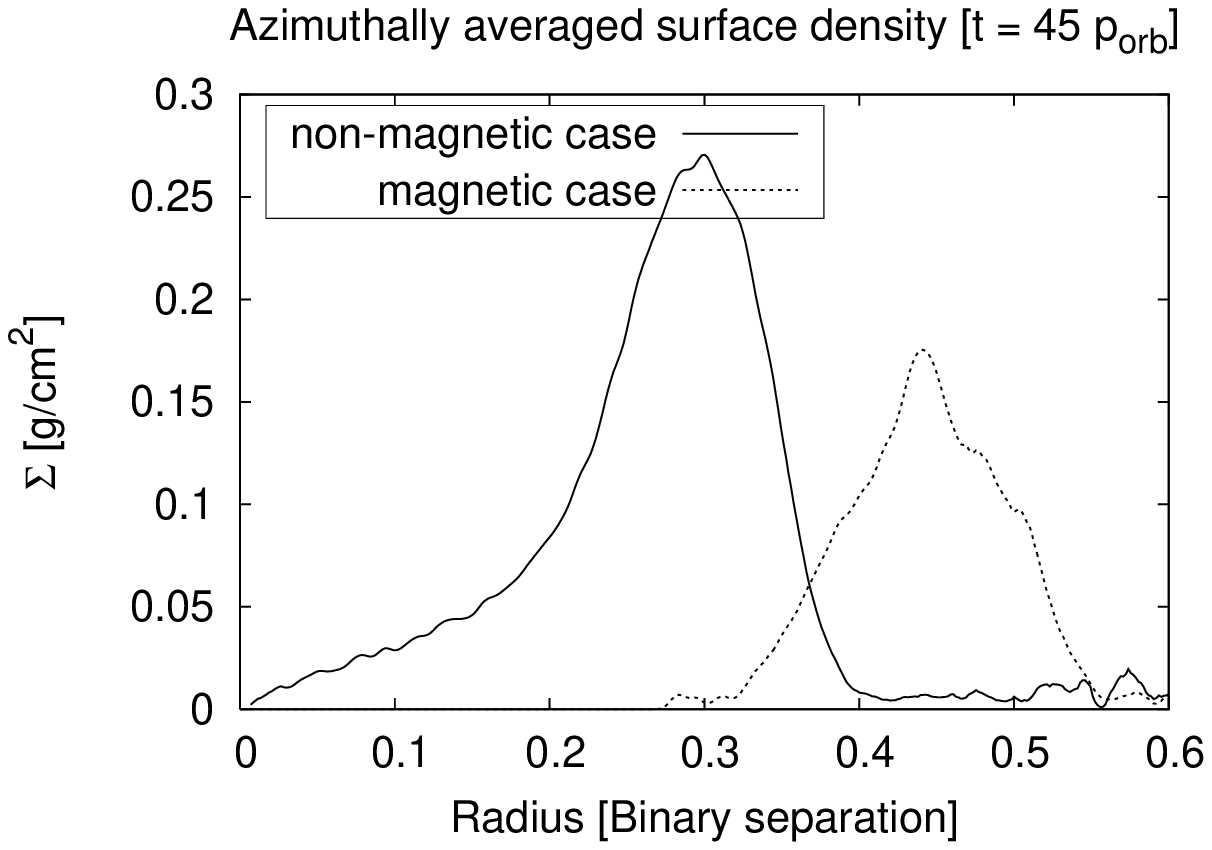}}}
\resizebox{80.0mm}{60.0mm}{
\mbox{
\includegraphics{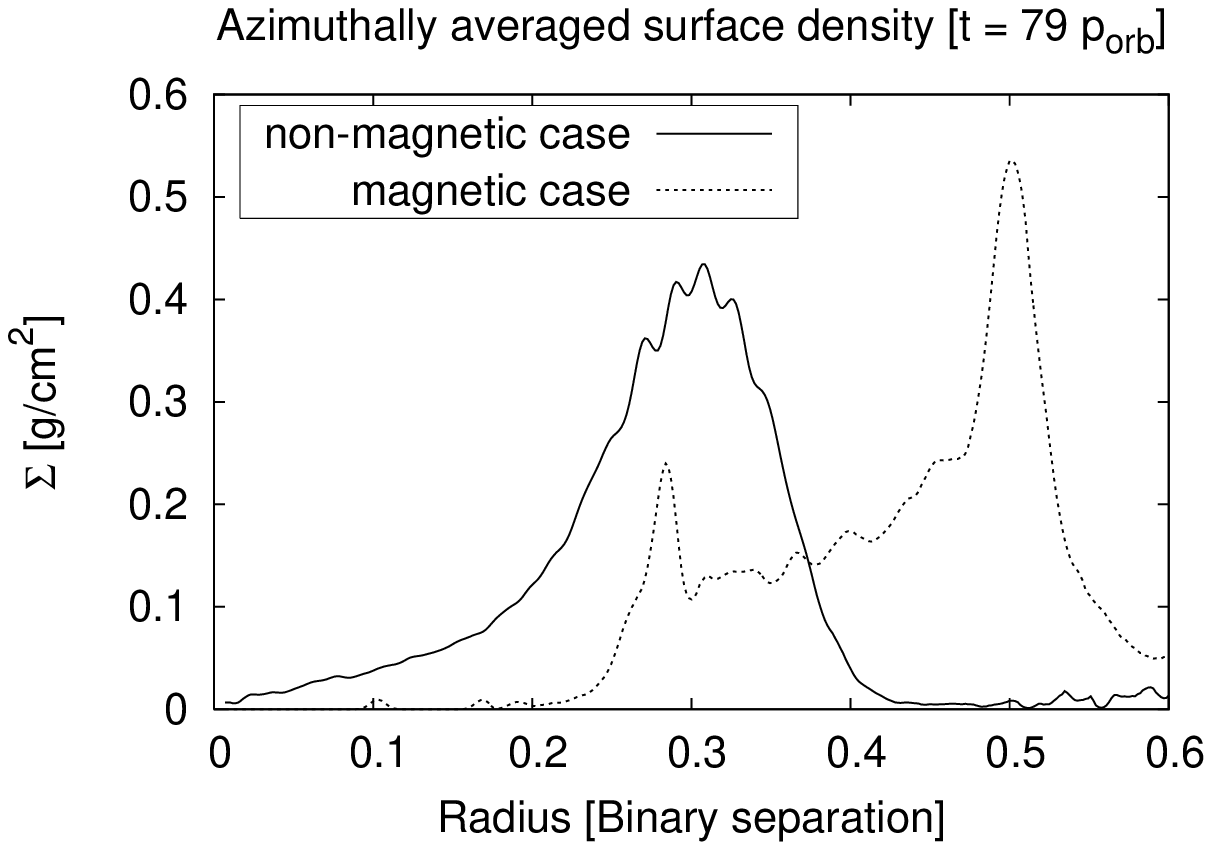}}}
\end{center}
\caption{
Azimuthally averaged plots of surface density against radius in two dimensional simulations of WZ Sge. The first plot shows the disc profile in the magnetic and non-magnetic cases after 45 orbits, before the disc becomes eccentric. The second plot shows the disc after 79 orbits, when the disc has become eccentric in the magnetic case. Note that the $\Sigma$ axis scale is not the same in both plots.}
\label{fig:discprofile3d}
\end{figure*}

\section{A magnetic solution?}
\label{sec:mag}

\subsection{WZ Sge as a magnetic propeller}
In sections \ref{sec:qdm} and \ref{sec:ext} we have argued that the value of
$\alpha_{\rm cold}$ in WZ~Sge may not be different from that in other
DN. This is consistent with the observed $\Delta M _{\rm acc}$, if
$R_{\rm out} \sim 0.5a$, and with the decay time of outbursts.  
However the recurrence time problem {\bf (Smak 1)} still remains unresolved. 

The observation of a 27.87 oscillation in the quiescent x-ray lightcurve
of WZ~Sge \citep{pat98} suggests that the white dwarf is magnetic.
The introduction of a magnetic field to a rapidly rotating white dwarf
raises the possibility of increasing the 
recurrence time by removing the inner, most unstable regions of the
accretion disc. This is a result of the magnetic torque 
which can, rather than accreting the inner disc, force mass close to the white 
dwarf to larger radii. This magnetic propeller effect occurs when the stellar 
magnetic field moves more rapidly than the disc material so that the disc 
gains angular momentum at the expense of the star.

\begin{figure}
\begin{center}
\resizebox{80.0mm}{60.0mm}{
\mbox{
\includegraphics{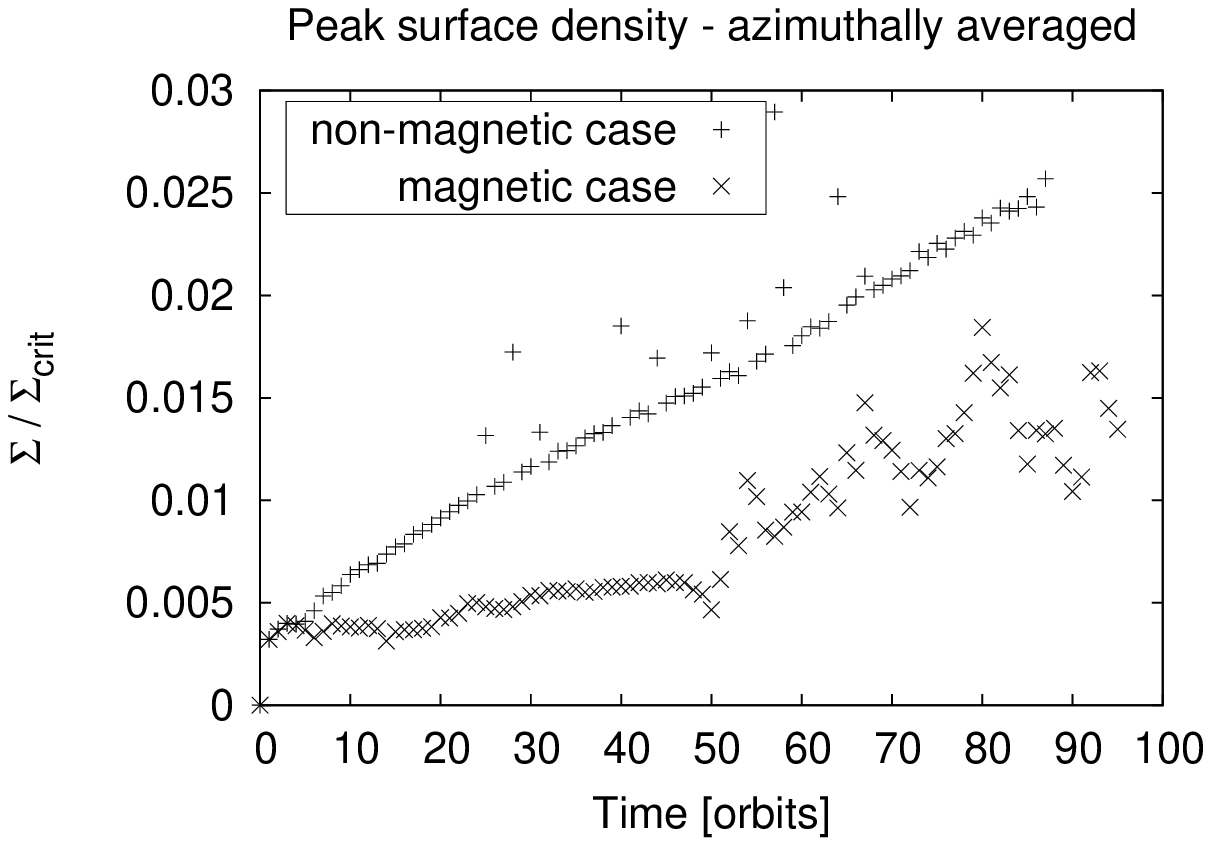}}}
\end{center}
\caption{Evolution of the peak surface density in the disc of WZ~Sge as a fraction of the critical surface density in non-magnetic and magnetic cases as a function of time. Results are taken from two-dimensional SPH calculations.
}
\label{fig:maggrownon}
\end{figure}

The inner disc can be depleted by the propeller and 
the surface density profile of the disc will then be altered as
material is forced to orbit further out in the disc, at radii close to the tidal
limit. The magnetic propeller inhibits accretion so that mass is forced to build up in
the outer disc, where the critical surface density is higher and the $\Sigma$ evolution time-scale ($\Sigma_{\rm crit}/\dot{\Sigma}$) is longer. The magnetic propeller 
effect cannot therefore only increase $t_{\rm rec}$ but can also allow the disc to 
accumulate enough mass prior to
outburst to supply an increased $\Delta {M_{\rm acc}}$, with the outer disc acting as a 
large reservoir of mass. In this 
way both {\bf (Smak 1)} and {\bf (Smak 2)} can be explained
simultaneously with a standard value for $\alpha_{\rm cold}$,
and without appealing to episodes of enhanced mass transfer. 

It should be noted that disc truncation is not normally, in itself, sufficient to increase 
recurrence times.  \citet{men00} show this by considering the evaporation 
of the inner disc. The fundamental difference between the magnetic propeller and 
other mechanisms such as evaporation and magnetically enhanced accretion is that the
propeller causes a build-up of mass in the outer disc by preventing accretion onto the 
white dwarf (during quiescence only). The overall profile of the disc is therefore 
completely altered, permitting much more mass to be stored in the disc. The mechanism 
presented here is analogous to that which was proposed by \citet{mat04} to explain long
recurrence times in FU~Ori type young stellar objects. 

The magnetic propeller should, in forcing the disc 
outwards, aid access to the 3:1 resonance. It is likely therefore that further mass will
be stored in the outer disc, as proposed by \citet{osa95}, providing an even larger
reservoir $\Delta M$ for mass between outbursts. This tidal effect may therefore
amplify the result of the weak propeller driving recurrence times to become even longer.

In the remainder of this 
section we outline the theoretical basis for the magnetic propeller model, and present 
the results of numerical experiments.

\subsection{The magnetic model}
\label{sec:tmm}
We adopt a description for the magnetic interaction which
assumes that as material moves through the magnetosphere it interacts with the
local magnetic field via a velocity dependent acceleration of the general form:
\begin{equation}\label{drag}
{a_{\rm mag}}=-k \left( \mathbf{v}-\mathbf{v}_{\rm f} \right)_{\bot}
\end{equation}  
where $\mathbf{v}$ and  $\mathbf{v}_{f}$ are the velocities
of the material and magnetic field respectively, 
and the suffix $\bot$ refers to the velocity components
perpendicular to the field lines. A version of this
description has been successfully applied to a number of other
intermediate polars by \citet{kin93}; \citet{wyn95}; \citet{wyn97} and 
\citet{kin99}. It has also been applied to discs around young stars by \citet{mat04} and \citet{mat04b}
The magnetic acceleration in equation (\ref{drag}) is intended to represent
the dominant term of the magnetic
interaction, with $k$ playing the role of a `magnetic $\alpha$'. The magnetic
time-scale can be written in terms of $k$ as
\begin{equation}\label{tmag}
{t_{\rm mag}}=k^{-1}\frac{|\mathbf{v}|}{|\mathbf{v}-\mathbf{v}_{\rm f}|_{\bot}}.
\label{eqn:tmaglast}
\end{equation}
It is assumed that the magnetic field of the WD is dipolar and that it rotates as a solid body with the star. We refer to a system as a weak magnetic propeller when,
in quiescence, the magnetic field is strong enough for the
system to behave as a magnetic propeller, but not strong enough
to eject a significant fraction of the transferred mass from the 
Roche lobe of the primary (as is the case in AE~Aqr). During outburst
the field should not be strong enough to prevent accretion.
This requires the following hierarchy of time-scales:
\begin{equation}
t_{\rm visc,h} \lesssim t_{\rm mag} \lesssim  t_{\rm visc,c}
\label{eqn:hier}
\end{equation}
where $t_{\rm visc,h}$ and $t_{\rm visc,c}$ are the hot and cold
viscous time-scales respectively. Assuming the
temperature of the quiescent disc to be $T \sim 5000$ K 
and $\alpha_{\rm c} =0.01$, and that in outburst we have 
$T \sim 5 \times 10^4$ K and $\alpha_{h}\rm{=0.1}$, equation (\ref{tvisc}) gives
\begin{equation} \label{cons}
1 \ {\rm days} \; \lesssim t_{\rm mag} \lesssim 40 \ {\rm days}.
\label{eqn:times}
\end{equation}
A further condition for a system to act as a weak propeller 
is that the magnetic interaction takes place at radii
exceeding the co-rotation radius $R_{\rm co} = (GM_1 P_{\rm spin}^2/4\pi^2)^{1/2}$,
where the fieldlines rotate with the local (circular) Kepler velocity.
A spin period of $27.87 \ {\rm s}$ for the WD in WZ~Sge gives
$R_{\rm co} \gtrsim R_{\rm WD} \sim 10^9$ cm, easily fulfilling this
requirement. If, conversely, the interaction were to take place within $R_{\rm co}$ then the effect of the magnetic field  would be to enhance accretion.

It now remains to quantify $t_{\rm mag}$. There are a
number of different models for the interaction of the inner disc and the white dwarf magnetosphere which have been applied to magnetic CVs.
Here we examine the two extreme cases in which the accretion disc
is completely magnetized, and conversely, in which the disc is diamagnetic. In both cases we still assume that the 
stellar field has a dipolar geometry, and that any local 
field distortions may be treated as perturbations on this structure. 

When the magnetic field lines permeate the inner
disc any vertical velocity shear  
tends to amplify the toroidal field component $B_{\varphi}$ \citep[e.g.][]{yi97}. 
Diffusive losses counteract this amplification and in
steady state, these effects balance to give 
\begin{equation}\label{tor}
\left | \frac{B_{\varphi}}{B_{\rm z}} \right | = \left | \frac{\gamma \left (
\Omega_{\star} - \Omega_{\rm K} \right )}{\tau \Omega_{\rm K}} \right |.
\end{equation}
Here $\gamma$ is a parameter that accounts for the uncertainty in the
vertical velocity shear and is of the order unity,
$\tau \la 1$ represents an uncertainty in the diffusive loss
time-scale, whilst $\Omega_{\star}$
and $\Omega_{\rm K}$ are the angular velocities of the WD and disc respectively.
The tension in the field lines results in a magnetic acceleration which may be
approximated by 
\begin{equation}\label{str}
a_{\rm mag} \simeq \frac{1}{\rho_{\rm d}{r_c}} \left ( \frac{{B_{\rm z}}^{2}}{4\pi} \right ) \hat{\mathbf{n}} \; ,
\end{equation}
where $\rho_{\rm d}$ is the disc density, 
$\hat{\mathbf{n}}$ is the unit vector perpendicular to
the field lines and $r_c$ is the radius of curvature of the field lines. 
It is possible to parametrize $r_c$ as\\
\begin{equation}\label{rad}
r_c \sim \epsilon{H}\left (\frac{B_{\rm z}}{B_{\varphi}} \right ) \; ,
\end{equation}
where $\epsilon$ is a constant of order unity.
Combining equations (\ref{tor}), (\ref{rad}) and (\ref{str}), with $H\sim\ {0.1R}$, 
the magnetic acceleration is given by
\begin{equation}
\label{eqn:amag}
a_{\rm mag} \sim - \frac{{5}\gamma{{B_{\rm z}}^{2}}}{2\pi R
\tau\epsilon\rho_{\rm d} {v}_{\rm K}}
\left( \mathbf{v}_{\rm K} - \mathbf{v}_{\rm f}\right)_{\bot} \; ,
\end{equation}
and $t_{\rm mag}$ by
\begin{equation}
\label{eqn:atime}
t_{\rm mag}\sim
\frac{2\pi{R}\tau\epsilon\rho_{\rm d} v_{\rm K}}{5\gamma {{B_{\rm
z}}^{2}}}\frac{\left| {v}_{\rm K} \right| }
{\left |{v}_{\rm K}-{{v}_{\rm f}}\right |_\bot} \; .
\end{equation}
At the circularisation radius $R_{\rm circ}\sim\rm{10^{10} \ cm}$, the disc density is
$\rho_{\rm d}\sim\rm{10^{-6}}\rm{\ g \ cm^{\rm{-3}}}$ (from equation (\ref{sig}), with
$H\sim\ {0.1R}$). Assuming $\tau\sim\epsilon\sim\gamma\sim\rm{1}$ 
we are able to constrain the magnetic moment
$\mu \sim B_{\rm z}R^{3}$ using (\ref{cons}), giving
\begin{equation}
2 \times \rm{10^{32} \ G\: cm^{3}} \lesssim \mu \lesssim \rm \rm{10^{33} \ G\: cm^{3}}
\end{equation}

For the opposite case of a diamagnetic flow we assume that the 
inner regions of the disc are broken up into a series of individual 
filaments, as predicated by the analysis of \citet{aly90}.
In order to follow the evolution of the diamagnetic gas blobs 
(which are formed well outside the co-rotation radius)  
we follow the approach of \citet{wyn97}. 
This approach assumes that  
the diamagnetic blobs interact with the field via a
surface drag force, characterized by the time-scale:
\begin{equation} 
{t_{\rm mag}}\sim{c_{\rm A}}\rho_{\rm b}l_{\rm b}B^{-2}\frac{|v_{\rm K}|}{|v_{\rm K}-v_{f}|_{\bot}}
\end{equation}
where $c_{\rm A}$ is the Alfv\'{e}n speed in the medium surrounding the plasma, $B$
is the local magnetic field, $\rho_{\rm b}$ is the plasma density, and ${l_{\rm b}}$ is
the typical length scale over which field lines are distorted. Making
the approximations that $\rho_b l_b \sim \Sigma_{\rm crit}(10^{10}
{\rm cm})$, $c_{\rm A} \simeq c_{\rm s}$ and ${\rm v} \simeq {\rm v}_{\rm K}$ we find that 
\begin{equation}
\rm{5 \times 10^{30} \ G\: cm^{3}} \lesssim \mu \lesssim \rm{3 \times 10^{31} \ G\: cm^{3}}.
\end{equation}

\subsection{One-dimensional numerical results}
\label{sec:cod}
\label{sec:res}
In this section we use a one-dimensional numerical model to confirm that it is feasible to create a depleted region in the centre of a quiescent accretion disc with a propeller. We also show that the propeller can lead to a build up of mass in the outer disc, and to extended recurrence times. The code outlined below is based on that used to similar effect in \citet{mat04}, but with the addition of viscosity switching to reproduce the effect of disc outbursts. A thin axisymmetric accretion disc is assumed to be rotating about a central mass $M$. The hydrodynamical equations are averaged over the azimuthal dimension so that non-axisymmetric effects, such as tidal forces from the secondary star, are not modelled. It is assumed that the disc is sufficiently cold that the dynamical time-scale $t_{\rm{dyn}}\sim R/v_\varphi$ is much smaller than the viscous or magnetic timescales so that a Keplerian approximation can then be adopted for the azimuthal motion. The magnetic torque due to the rotating WD is added using a parametrization of the form
\begin{equation}\label{torque}
\Lambd = \frac{l}{t_{\Lambd}} = \frac{\sqrt{G M}R^{1/2}}{t_\Lambd}
\; ,
\end{equation}
where $G$ represents the universal constant of gravitation, $M$ is the mass of the central star and $t_\Lambd$ is the time-scale on which the local disc material
gains angular momentum, which in this case is given by equation (\ref{eqn:atime}). The resulting equation for the evolution of surface density with time \citep[derived in][]{mat04} is
\begin{eqnarray}\label{sigmaevol}
\lefteqn{\frac{\partial\Sig}{\partial t} =
{} \frac{3}{R}\frac{\partial}{\partial R}\left[
R^{1/2}
\frac{\partial(R^{1/2}\nue\Sig)}{\partial R}
\right]}
\\ 
\nonumber
 & &
{}- 
\frac{{\it \beta}}{R}\frac{\partial}{\partial R}\left[
\frac{1}{R^{7/2}}
\left(\left[\frac{R}{R_\mathrm{co}}\right]^{3/2} - 1\right)
\right]
\; ,
\nonumber
\end{eqnarray}
where $\beta$ is defined as
\begin{equation}
\label{betalast}
\beta \sim \frac{\mbox{\boldmath{${\mu}$}}^{2}}{2 \pi \sqrt{G M}} 
\; .
\end{equation}
We adopt the \citet{sha73} viscosity prescription (\ref{eqn:shasun}) and the viscous dissipation is calculated using equation (\ref{eqn:emission}). The thermal viscous disc instability (DIM) must be modelled if disc outburst behaviour is to be reproduced, and the viscosity is therefore permitted also to vary in time. The approach to viscosity switching adopted here is very similar to that used by \citet{tru00} in their smoothed particle hydrodynamic simulations. Disc annuli are switched between high and low viscosity states, by altering the viscous $\alpha$ parameter, using critical density `triggers' \citep{can88} which lie at the extremes of the limit cycle. Accordingly when the local surface density $\Sigma$ exceeds that of equation (\ref{sig}) the local viscous parameter is altered smoothly on a thermal timescale ($t_{\rm th} \sim \left(\alpha \Omega \right)^{-1}$) to a value of $\alpha_{\rm hot} = 10 \alpha_{\rm cold} = 0.1$. Later, when $\Sigma$ falls below a lower trigger density, the viscous parameter is returned to its original value $\alpha_{\rm cold} = 0.01$. In the high state the accretion disc should also be hotter, so the temperature $T$ is switched between $5000 \ {\rm K}$ and $50000 \ {\rm K}$ in the same way.

The code calculates the diffusive and advective terms separately on a grid of resolution 500 and combines the results using operator splitting. The diffusive term is calculated using a forward in time space-centred scheme (FTSC) and the values of $\Sig$ produced here are used as input for a modified two-step Lax-Wendroff scheme \citep{pre92} which is used to solve the advective term. The inner boundary, representing the stellar boundary layer, is placed at $R_{\star} = 1.0 \times 10^{9} \ {\rm cm}$ and an outflow condition such that $\frac{\partial ^{2} \Sig}{\partial R^{2}}$ is constant is applied. An inflow condition is applied at the outer boundary which is placed at the $0.4a$, following SR. The stellar mass is set to $M = 1.2 \ \msun$ and mass is added at the outer boundary at a rate of $\mdot = 2 \times 10^{-10} \ \msun \ {\rm yr}^{-1}$. The white dwarf spin period is set to $P_{\rm spin} = 28 \ {\rm s}$ which gives a co-rotation radius of $R_{\rm co} = 1.5 \times 10^{9} \ {\rm cm}$. 

Calculations were performed without a primary magnetic field and with a field of $B_{z} \sim 1 \ {\rm kG}$ ($\mu \sim 10^{30} \ {\rm G cm^{3}}$), although the precise value depends on the prescription adopted for the magnetic acceleration. The effect of the magnetic field was to truncate the inner disc, producing a depleted central region and to reduce the peak surface density.  The inner disc is likely to be the starting point of any outburst in the non-magnetic case while in the magnetically truncated case this should instead occur close to the truncation radius. It is important that in neither case is the disc close to marginal stability, where the steady state profile would peak very close to $\Sigma/\Sigma_{\rm crit}=1$. Such marginal stability is another possible reason for long outburst times, but it would not cause an increase in outburst amplitude.

When the viscosity switching mechanism was enabled a series of outbursts occurred in both the magnetic and non-magnetic cases. These are plotted in Fig \ref{fig:newcurve}. In the non-magnetic case the outburst time is $t_{\rm ob} \sim 10 \ {\rm d}$ and the recurrence time is $\trec \sim 20 \ {\rm d}$. This is roughly in agreement with observations of normal DN, although the outbursts are longer than expected. The outbursts vary in amplitude but are of the expected order of magnitude. In the magnetic case the recurrence time is increased by $\sim 40$ times to $\trec \sim 2 \ {\rm yr}$, and the outburst amplitude is increased by a factor of $\sim 2$. In both cases the quiescent dissipation is more than two orders of magnitude less than the outburst dissipation. The profile of an individual magnetically moderated outburst is shown in Fig \ref{fig:obprof}. The outburst lasts for $\tob \sim 10 \ {\rm d}$, notably shorter than is observed. It appears, after an initial sharp rise, to rise on a time-scale of a few days. This is close to the hot viscous timescale, rather than to the thermal time-scale. This is probably because once the disc becomes hot, on a thermal time-scale, mass migrates into the centre of the disc, filling the evacuated central regions on the viscous timescale. The outburst decays on a similar viscous time-scale, and then much more quickly, on the thermal time-scale, as the disc returns to the cool state.

Fig \ref{fig:massevolve} shows how the mass of the accretion disc varied during outbursts. The total peak disc mass in the magnetic case was $M_{\rm disc} \sim 8 \times 10^{23} \ {\rm g}$, twice that of the non-magnetic case. Further, the mass consumed in the magnetic outbursts was $\Delta \sim 4 \times 10^{23} \ {\rm g}$, ten times that consumed in the non-magnetic outbursts, half of the total accretion disc mass, and approaching the value deduced from observations. We also note that the mass growth rate in the magnetic case is close to linear, this confirms that the disc is not close to marginal stability, as was discussed above. The results from the non-magnetic case show a series of alternate outbursts and mini-outbursts. Mass accretion onto the white dwarf is plotted in Fig \ref{fig:mdot}, on the same time axis as Fig \ref{fig:massevolve}. The accretion rate peaks at $5 \times 10^{-9} \ {\rm \msun \ yr^{-1}}$ in the magnetic case, which is a factor of $2$ more than the non-magnetic case. It is also notable that the cold viscous decay, which is not visible in Fig \ref{fig:obprof}, can be clearly seen in the magnetic case.  

In Fig \ref{fig:newevol} the detailed evolution of a magnetically moderated outburst can be followed. During quiescence the propeller has prevented the disc from spreading in to the star, and by preventing accretion has forced a large build up of mass in the outer disc. The outburst begins when the surface density first exceeds the critical value, close to the truncation radius, rather than near the boundary layer in the usual way. Heating waves then sweep inwards and outwards on a thermal time-scale until the whole disc is in the hot state. The viscosity of the disc is now higher, and the viscous timescale shorter, allowing mass to spread inwards and quench the propeller. The disc now takes a form close to the standard non-truncated Shakura-Sunyaev solution. Mass accretion onto the white dwarf is high and the density of the disc drops rapidly until the lower trigger is reached at the outer edge of the disc. At this point a cooling wave sweeps inwards so that the disc becomes cool again and is gradually shifted outwards as the propeller is reestablished. Around half of the total disc mass is consumed in the outburst.

Results from these calculations confirm that a magnetic propeller is capable of truncating the inner accretion disc and may also lead to an extended recurrence time. However the recurrence time is still an order of magnitude shorter than that observed with these simulation parameters. Two possible mechanisms for further extending the recurrence time suggest themselves. The first requires only a slightly stronger magnetic field. If the magnetic field is stronger then the disc will be truncated at a larger radius, and this can be tuned so that the disc is very close to the steady state when it reaches the outburst trigger. As the disc approaches this steady state its growth rate slows asymptotically to zero so that the recurrence time in this case can be extremely long. This effect was discussed by \citet{war96,men00}. It is distinct from, and acts in addition to the mechanism described above which is concerned instead with the linear regime of disc growth. Results from a calculation using this marginal stability effect are shown Fig. \ref{fig:marginal}. The recurrence time is increased to $t_{\rm rec} \sim 20 \ {\rm yr}$ but neither the mass consumed nor the amplitude of the outbursts is greatly altered from the standard magnetic case illustrated in Fig \ref{fig:newcurve}, so that the marginal stability effect would be unable, in itself, to solve {\bf (Smak 2)}. This mechanism also requires fine tuning; if the field is slightly too strong outbursts will be prevented entirely. 

A second possible explanation for the further extension of the recurrence time, which does not require such fine tuning, is that the size of the mass reservoir  may be further increased by tidal effects. The one-dimensional treatment detailed above precludes the consideration of binary tidal forces and resonances. To better model the structure of the accretion disc, including its non-axisymmetric modes, we turn to the complementary two-dimensional model described in section \ref{sec:threed}.

\subsection{Two-dimensional numerical results}
\label{sec:threed}
Further numerical calculations were performed in two dimensions with smoothed particle hydrodynamics (SPH) simulations \citep[e.g.][]{kun97,sch04}. The above prescription (\ref{eqn:amag}) for the magnetic acceleration was implemented in an SPH code and calculations were performed with and without a dipolar magnetic field in the plane of the accretion disc. The system parameters were similar those used in the one-dimensional simulations described above, although the mass transfer rate was $\mdot = 3 \times 10^{-11} \ \msun \ {\rm yr^{-1}}$. Each calculation used around 40,000 particles with a fixed smoothing length. Particles were injected at the first Lagrangian point and removed when they approached the primary star. 

The resulting discs are shown in Figs \ref{fig:disc} and \ref{fig:discprofile3d}, plotted as face-on disc views and radial profiles respectively. The profiles are somewhat different to those from the one-dimensional simulations, shown in Fig \ref{fig:newevol}, as a result of the torques from the binary potential. The effect of the magnetic field in pushing the disc outwards is more pronounced than in the one dimensional model. In the magnetic case, because matter is driven towards the resonant radii, the disc becomes eccentric and begins to develop some spiral structure. These are clearly seen as two peaks in the second plot of Fig \ref{fig:discprofile3d}. 

This structure also affects the time evolution of the surface density. Fig \ref{fig:maggrownon} shows how the peak surface density as a fraction of the critical surface density grows with time in the magnetic and non-magnetic cases. This can be viewed as a measure of how close the disc is to outburst, since when any substantial part of the disc reaches its local critical density an outburst will be triggered. In the non-magnetic case there is a simple monotonic growth of this parameter. The noise visible in the otherwise very smooth trend is caused by density enhancements associated with accretion events at the inner boundary, which may be a result of the limited resolution of the simulation. This monotonic growth is as expected from Fig \ref{fig:discprofile3d} which shows that the non-magnetic disc profile keeps the same form, with the density peak remaining at the same point around $\sim 0.3 a$. If surface density continues to grow at this rate until outburst then this simulation represents around 3\% of the duty cycle for a non-magnetic system. In the magnetic case the same parameter grows monotonically at first, much more slowly than in the non-magnetic case. Its behaviour becomes stochastic once the disc is driven to become eccentric. It is problematic to extrapolate from such a short period to the length of the recurrence time, however the overall trend remains clear: the approach to outburst is {\it clearly slower} in the magnetic case. This result does not contradict that of \ref{sec:cod}. It therefore seems unlikely that effects due to the binary would prevent the propeller from leading to a long recurrence time. It also remains possible that the tidal forces could lead to the development of a much larger reservoir, and hence a longer recurrence time, than in the simple axisymmetric case.

\section{Observational consequences}
\label{sec:obs}
The weak magnetic propeller model for the outbursts of WZ~Sge results
in a number of observationally testable predictions. The major
consequence of the model, the existence of a substantial hole in the 
inner disc, has already been suggested by \citet{men98}.
The authors estimate the ratio of the inner and outer disc radii
($R = R_{\rm in}/R_{\rm out}$) from observations of the H$_\alpha$
emission line widths. WZ~Sge was found to show an extremely large 
value of $R \sim 0.3$. This value agrees reasonably well with the second plot in Fig
\ref{fig:disc} 
where $R_{\rm out} \simeq 0.5 a$ and $R_{\rm in} \simeq 0.2 a$ for the magnetic case, and
is much larger than the value expected if the accretion disc extended
all the way to the WD surface ($R \sim 0.03$). 

The asynchronous magnetic cataclysmic variables (the intermediate
polars) show multiple periodicities in their 
light curves. WZ~Sge has been observed to display oscillations at 
27.87~s and 28.96~s as well as weaker oscillations at 
28.2~s and 29.69~s \citep[][and references therein]{las99}. \citet{las99} 
interpret WZ Sge as an intermediate polar, and identify the 27.87~s 
oscillation as arising from the spinning white dwarf, and the other
oscillations as beat periods between the white dwarf
spin and material orbiting within the disc. The beat periods arise
from the reprocessing of x-ray emission from the WD on the surface of
the disc, the hotspot and/or the secondary star.
Assuming the material in the disc is moving in circular, Keplerian orbits
the radius associated with these beat periods $\rm{R_{\rm beat}}$ can be written as
\begin{equation}
{R_{\rm beat}}=\left(\frac{{P_{\rm K}^{2}}GM}{4\pi^{\rm{2}}}\right)^{\rm{1/3}}
\end{equation}
where
\begin{equation}
\frac{1}{P_{\rm K}}=\frac{1}{P_{\rm spin}}-\frac{1}{P_{\rm obs}}.
\end{equation}
Here ${P_{\rm K}}$ is the Keplerian period of the disc material, and ${P_{\rm obs}}$ is the observed beat period.
Assuming $M= 1.2 {M}_{\odot}$ we find
$R_{28.96} \simeq 1.3 \times 10^{10}$ cm ($\simeq 0.3a$),
$R_{28.2} \simeq 2.8 \times 10^{10}$ cm ($\simeq 0.6a$), and
$R_{29.69} \simeq 9.4 \times 10^{9}$ cm ($\simeq 0.2a$).
The radii $R_{28.96}$ and $R_{29.69}$ correlate reasonably well with
material orbiting close to the inner edge of the quiescent disc in two-dimensional numerical models (cf Fig \ref{fig:disc}). The radius $R_{28.2}$ on the other hand,
is quite close to the hotspot (stream/disc impact region) which should be located at $\simeq 0.5a$, as seen in the magnetic cases in Fig \ref{fig:disc}. Obtaining a closer fit between these
radii and the model would require tighter constraints on the WD mass,
and $q$, as well a suitable model for the x-ray reprocessing within
the disc. Nonetheless absence of beat radii from the supposed evacuated region is encouraging for the magnetic propeller model. The fact that the oscillations are not observed during outburst is also entirely consistent with this model: the magnetosphere is compressed during outburst, and the system is observationally ``non-magnetic''. The model outlined in this paper predicts that WZ Sge has an eccentric precessing disc, and it might therefore be expected to display permanent superhumps. 

\section{The spin evolution of the white dwarf}
\label{sec:spi}
The spin evolution of the white dwarf in WZ Sge is determined by
angular momentum transfer in the quiescent and outburst phases. 
In quiescence the WD acts as a magnetic propeller, i.e. material 
approaching the WD is forced to larger radii. We can approximate this 
interaction by assuming the transferred gas, with initial specific angular
momentum $(G M R_{\rm circ})^{1/2}$, is forced to orbit at a mean 
outer radius $\bar{R}_{\rm out}$.  The associated spin down
torque on the WD is then 
\begin{equation} \label{jd}
\dot{J}_{\rm d} = - \zeta \dot{M}_2 {\left (GM \right )}^{1/2} 
\left(\bar{R}_{\rm out}^{1/2} - R_{\rm circ}^{1/2}\right),
\end{equation}
where 
the mass transfer rate accross the L1 point $\dot{M}_{2} \sim \rm{10^{15}gs^{-1}}$, $\bar{R}_{\rm out} \sim{0.5a}$, 
$R_{\rm circ} \sim \rm{0.35a}$, and
the parameter $\zeta > \rm{1}$ if material at $\bar{R}_{\rm out}$ loses
angular momentum to the secondary star via tides (this will be
expected to occur once the outer edge of the disc reaches the tidal radius). 
Conversely, in outburst, the WD gains angular momentum with accreted matter at the rate
\begin{equation} \label{ju}
\dot{J}_{\rm u} ={\dot{M}_{\rm acc}}(GM R_{\rm in})^{1/2} \; ,
\end{equation}
where we adopt an inner disc radius of $R_{\rm in} = 10^{9} \ {\rm cm}$. From the results of \citet{sma93} the average accretion rate during outburst can be estimated as
${\dot{M}_{\rm acc}} \sim  10^{18} \ {\rm g \ s^{-1}}$. \\
The evolution time-scale of the white dwarf spin period 
can be written in terms of $\dot{J}$ as
\begin{equation}
\tau_{\rm spin} \sim \frac{P_{\rm spin}}{\dot{P}_{\rm spin}} = - \frac{2\pi I}{P_{\rm spin}\dot{J}},
\end{equation}
where $I$ is the moment of inertia of the WD. Averaging over a duty
cycle the torque on the WD is given by
\begin{equation}
<\dot{J}> \simeq \frac{\dot{J}_{\rm u} t_{\rm ob} - \dot{J}_{\rm d} t_{\rm rec}}
{(t_{\rm rec} + t_{\rm o})}
\end{equation}
where $t_{\rm ob} \sim 50 \ {\rm d}$ is the outburst duration and we take $t_{\rm rec} \sim 30 \ {\rm yr}$. From the estimates above
(assuming $\zeta \sim 1$)
we find $\dot{J}_{\rm u} t_{\rm o} / \dot{J}_{\rm d} t_{\rm rec} \sim 6$. The greater spin-up torque gives a spin up on the time-scale
\begin{equation}
\tau_{\rm d}\sim \frac{P_{\rm spin}}{|\dot{P}_{\rm spin}|} 
\simeq  -\frac{2\pi I}{P_{\rm spin}|<\dot{J}>|} 
\sim 10^{9} \ {\rm yr}.
\end{equation}
The WD in a weak propeller may experience a net spin up, as above, or a net spin down, if outbursts are shorter or the mass transfer rate is higher for example. Both these results may lead to a spin up/spin down cycle.\\

\noindent {\bf Net spin up:} In this case, illustrated by WZ~Sge, the net spin evolution is towards shorter periods. Therefore, according to equation (\ref{drag}) the propeller will become stronger with time. The condition $t_{\rm mag} < t_{\rm visc,h}$ will eventually be satisfied. At this point accretion will be prevented completely and outbursts will cease; the system will become AE~Aqr like. This phase should be short lived however and it must lead to a spin down. Therefore a cycle between WZ~Sge and AE~Aqr type stars may arise.\\

\noindent {\bf Net spin down:} In this opposite case, the net evolution is towards longer white dwarf spin periods. The propeller becomes weaker with time until it is eventually quenched when $t_{\rm mag} > t_{\rm visc,c}$. The system would then look like a normal SU~UMa star. This SU~UMa phase would usually result in white dwarf spin up. Therefore an alternative cycle may exist between WZ~Sge and SU~UMa type systems. Both of these cycles are a natural consequence of hierarchy (\ref{eqn:hier}).

%The slightly larger spin-up torque leads therefore to the prediction that the 
%WD spin period is currently increasing on the time-scale
%\begin{equation}
%\tau_{\rm d}\sim \frac{P_{\rm spin}}{|\dot{P}_{\rm spin}|} 
%\simeq  -\frac{2\pi I}{P_{\rm spin}|<\dot{J}>|} 
%\sim \rm{4}\times 10^{8} \ {\rm yr}.
%\end{equation}
%An increasing $P_{\rm spin}$ will eventually lead to the condition 
%$t_{\rm mag} > t_{\rm visc,c}$ and
%the propeller will be quenched, allowing accretion to take place 
%during quiescence. If we assume $t_{\rm mag} \sim 50 \ {\rm d}$ at present and use a magnetic interaction radius of $10^{10} \ {\rm cm}$ then equation (\ref{drag}) suggests that the propeller will be quenched when $P_{\rm spin} \sim 50 \ {\rm s}$.
%Once the WD has spun down to the required, period accretion will begin,
%spinning up the WD. Hence the long term evolution of the WD spin may be a cycle of spin down (weak propeller) and spin up (standard
%accretion disc) phases with the full cycle time being of order 
%$10^9$ years. This is a natural consequence of hierarchy (\ref{eqn:hier}),
%and depends only upon the white dwarf being sufficiently magnetic.
%The main observational characteristic of the weak propeller phases
%is the very long inter-outburst time-scale. During the spin up phases
%WZ~Sge would be observationally indistinguishable from other ``normal''
%SU~UMa stars with similar orbital periods.

\section{Discussion}
\label{sec:dis}
We have argued that the outbursts of WZ Sge can be explained using 
the standard DIM value for $\alpha_{\rm cold}$ if we assume that the 
white dwarf in the system acts as a magnetic
propeller. Numerical models of the magnetic disc, in both one and two dimensions, predict an increased inter-outburst time $t_{\rm rec}$,
and a disc massive enough 
to fuel an increased outburst accretion rate. No episode of enhanced mass
transfer from the secondary star is required to trigger outbursts, or to supply mass
during outburst. 

The weak-propeller models assume that the
magnetic tension force is the dominant cause of angular momentum
transfer in the quiescent disc, and that the force is proportional to the 
local shear between the disc plasma and the magnetic field. In
outburst angular momentum transfer is dominated by viscous diffusion, and
mass accretion on to the white dwarf is permitted.  This results in an 
estimated WD magnetic moment in the range
\begin{equation}
2 \times 10^{32} \ {\rm G \ cm^3} \lesssim \mu \lesssim 10^{33} \ {\rm G \ cm^3} 
\end{equation}
if the disc is assumed to be fully magnetized, or a value in the range
\begin{equation}
5 \times 10^{30} \ {\rm G \ cm^3} \lesssim \mu \lesssim 3 \times 10^{31} \ {\rm G \ cm^3} 
\end{equation}
if the disc is assumed to be diamagnetic. These values for the WD
magnetic moment would suggest that WZ~Sge is a short period
equivalent of the intermediate polars (IPs). Most IPs lie above 
period gap ($P_{\rm orb} \gtrsim 3 \ {\rm hours}$) and have magnetic moments in
the range $10^{31} \ {\rm G \ cm^3} \lesssim \mu \lesssim 10^{34} \ {\rm G \ cm^{3}}$. \citet{kin99} pointed out the only two confirmed IPs with 
$P_{\rm orb} \lesssim 2$ hours (EX~Hya and RX1238-38) have $\mu \gtrsim
10^{33}$ G cm$^3$ and long spin periods (67 min and 36 min
respectively). WZ~Sge would therefore be the first weakly magnetic $\mu
< 10^{32} \ {\rm G \ cm^{3}}$ CV to be found below the period gap. 

In differentiating between
the models of the disc-magnetosphere interaction it is useful to 
consider the case of AE~Aquarii. This system is thought to contain a WD with 
a magnetic moment $\sim 10^{32} \ {\rm G \ cm{^3}}$, which ejects $\sim$ 99\% 
of the mass transfer stream from the system \citep{wyn97}.
If the WD magnetic moment $\mu \gtrsim
10^{31}$ in WZ~Sge then it would satisfy the criteria necessary to become an
ejector system. Hence in the case a fully magnetized disc, 
it would seem that the presence of the disc itself 
is the only protection the system has from ejecting the transferred mass. 
If the disc was completely accreted or destroyed for any reason 
WZ~Sge would resemble a short period version of AE~Aqr. 
This would not apply to the case where the WD has a magnetic moment in 
the range predicted by the diamagnetic model. 

A WD spin evolution cycle results as a natural consequence 
of the time-scale hierarchy (\ref{eqn:hier}). In the case of WZ~Sge we expect the spin cycle to run between the current state, characterised by long inter-outburst times, and that of an AE~Aqr-like strong propeller. In some similar systems however, spin-up phases would look like a normal SU~UMa system, whilst the spin down phases would be WZ~Sge like. WZ~Sge-like systems following such spin 
cycles are expected to be relatively rare because of the constraint
(\ref{eqn:hier}), which results in the restricted range of allowed WD magnetic
moments presented above. 

\citet{men98} find evidence for ring-like accretion
discs in long supercycle length SU~UMa stars from a radial velocity
study of the $H_\alpha$ emission lines of these systems. WZ~Sge is
the most extreme of these objects with a recurrence time of $\sim 12,000 \ {\rm d}$
and $R = R_{\rm in}/R_{\rm out} = 0.3$, which is in good
agreement with our numerical results. Other SU~UMas
show a strong positive correlation between supercycle times and the
$R$ parameter. This result is in agreement with the model presented
in this paper for WZ~Sge. The WDs in the other, less extreme, SU~UMa 
stars would simply act as less efficient propellers: the WDs
would have weaker magnetic moments, or lower spin rates (which
may reflect various stages of the spin up/down cycles postulated above).
Strong candidates for WZ~Sge-like systems are RZ~Leo ($t_{\rm rec} <$
4259 days, $R \simeq 0.16$), CU~Vel ($t_{\rm rec} \simeq$  700 - 900
days, $R \simeq 0.15$) and WX~Cet ($t_{\rm rec} \simeq$ 1000 days, 
$R \simeq 0.12$), \citep[taken from][table 5]{men98}.

\section*{Acknowledgments}
We would like to thank Erik Kuulkers, James Murray, Joe Patterson and
Rudolf Stehle for help in preparing this manuscript. We also thank the referee, Jean-Marie Hameury, for his helpful review. Research in theoretical astrophysics at the University of Leicester is supported by a PPARC rolling grant.

\end{document}